\documentclass[iop,apj,tighten,numberedappendix]{emulateapj}
\usepackage[breaklinks,colorlinks,urlcolor=blue,citecolor=blue,linkcolor=blue]{hyperref}

\usepackage{amsmath,amssymb}
\usepackage{cleveref}
\usepackage{graphicx}
\usepackage{bm}
\usepackage[toc,title,page]{appendix}
\usepackage{tocbibind}
\usepackage{xcolor,ulem}
\usepackage{color}
\newcommand{\Msun}{{\rm M}_\odot}

\def\lsim{~\rlap{$<$}{\lower 1.0ex\hbox{$\sim$}}}
\def\gsim{~\rlap{$>$}{\lower 1.0ex\hbox{$\sim$}}}

\shorttitle{Emission for optical flares with GWs}
\shortauthors{Tagawa~et~al.}

\begin{document}
\title{
Shock cooling and breakout emission for 
optical flares associated with gravitational wave events
}

\author{
Hiromichi Tagawa\altaffilmark{1}, 
Shigeo S Kimura\altaffilmark{2,3},
Zolt\'an Haiman\altaffilmark{4,5},
Rosalba Perna\altaffilmark{6,7},
Imre Bartos\altaffilmark{8}
}
\affil{
\altaffilmark{1}Shanghai Astronomical Observatory, Shanghai, 200030, People$^{\prime}$s Republic of China\\
\altaffilmark{2}Astronomical Institute, Graduate School of Science, Tohoku University, Aoba, Sendai 980-8578, Japan\\
\altaffilmark{3}Frontier Research Institute for Interdisciplinary Sciences, Tohoku University, Sendai 980-8578, Japan\\
\altaffilmark{4}Department of Astronomy, Columbia University, 550 W. 120th St., New York, NY, 10027, USA\\
\altaffilmark{5}Department of Physics, Columbia University, 550 W. 120th St., New York, NY, 10027, USA\\
\altaffilmark{6}Department of Physics and Astronomy, Stony Brook University, Stony Brook, NY 11794-3800, USA\\
\altaffilmark{7}Center for Computational Astrophysics, Flatiron Institute, New York, NY 10010, USA\\
\altaffilmark{8}Department of Physics, University of Florida, PO Box 118440, Gainesville, FL 32611, USA\\
}
\email{E-mail: htagawa@shao.ac.cn}

\begin{abstract} 
The astrophysical origin of stellar-mass black hole (BH) mergers discovered through gravitational waves (GWs) 
is widely debated. 
Mergers in the disks of active galactic nuclei (AGN) represent promising environments for
at least a fraction of these events, 
with possible observational clues in the GW data. 
An additional clue to unveil AGN merger environments 
is provided by possible electromagnetic emission from 
post-merger 
accreting BHs. 
Associated with BH mergers in AGN disks, 
emission from shocks emerging around 
jets launched by accreting merger remnants is expected. 
In this paper
we compute the properties of the emission produced during breakout and the subsequent adiabatic expansion phase of the shocks, and  we then apply this model to optical flares 
suggested to be possibly
associated with GW events. 
We find that the majority of the reported flares 
can be explained by 
the breakout and the shock cooling emission. 
If these events are real, then the
merging locations of binaries 
are constrained depending on the emission processes. 
If the optical flares 
are produced by shock cooling emission, they would display 
moderate color evolution, 
possibly  color variations among different events, 
a positive correlation between the delay time and the duration of flares, 
and accompanying breakout emission in X-ray bands before the optical flares. 
If the breakout emission dominates the observed lightcurve, 
it is expected that 
the color is distributed in a narrow range in the optical band, and 
the delay time from GW to electromagnetic emission is longer than $\sim 2$ days. 
Hence, further explorations of the distributions of delay times, color evolution of the flares, and associated X-ray emission will be useful to test the 
proposed emission model for the observed flares.
\end{abstract}

\section{Introduction}

The astrophysical pathways to black hole (BH) mergers discovered by the LIGO \citep{2015CQGra..32g4001L}, Virgo \citep{2015CQGra..32b4001A}, and KAGRA \citep{2021PTEP.2021eA101A} 
gravitational wave (GW) observatories 
have been actively debated \citep{Barack2019}. 
Various scenarios have been proposed, including 
isolated binary evolution \citep[e.g.][]{Dominik12,Kinugawa14,Belczynski16,Spera19,Tagawa18,Tanikawa2022} 
evolution of triple or quadruple systems
\citep[e.g.][]{Silsbee17,Antonini17,Michaely19,Fragione19,Martinez2022,2023arXiv230210350B}, 
dynamical evolution in star clusters 
\citep[e.g.][]{PortegiesZwart00,Samsing14,OLeary16,Rodriguez16,Banerjee17,Kumamoto18,Rasskazov19,Perna2019,Fragione18_EvolvingGC,Antonini2022}, 
and compact objects in active galactic nucleus (AGN) disks
\citep[e.g.][]{Bartos17,Stone17,McKernan17,Tagawa19,ArcaSedda2023}.

In an AGN disk, 
BHs are embedded in due to capture via dynamical interactions between the nuclear star cluster and the AGN disk \citep{Ostriker1983,Miralda-Escude2005,Generozov2022,Wang2023_capture}
and by in-situ star formation \citep{Levin2003,Goodman04,Milosavljevic2004,Derdzinski23,ChenY2023}. 
The AGN disk environment 
helps to bring the BHs closer together
\citep{Bellovary16,Tagawa19,Derdzinski19,Derdzinski21,Li2021_BinaryEv,Grishin2023} and hence form binaries \citep{LiJiaru2022,Boekholt2023,Rowan2022,LiJiaru2023,DeLaurentiis2023,Rozner2023,Whitehead2023,Rowan2023,Qian2023} which then  merge over relatively short time scales. 
Comparisons to the observed BH masses, spins and merger rate indicate that a sizable fraction of the observed mergers may indeed originate in AGN disks \citep{2021ApJ...920L..42G,Tagawa20_MassGap,Ford2022}. 
The AGN channel could also explain some of the peculiar detections, such as those with a high mass \citep{Tagawa20_MassGap,Gayathri2023} and possibly high eccentricity (\citealt{Samsing20,Tagawa20_ecc,Gayathri2022,RomeroShaw2022} but see \citealt{Romero-Shaw20,RomeroShaw2023}). 

Due to the gas-rich merger environment, a key signature of the AGN channel is the possibility of electromagnetic emission accompanying the GW signal from the merger \citep{Bartos17,Stone17}. 
To explore this possibility, electromagnetic follow-up observations have been carried out for many of the mergers, with 
nine counterpart candidates suggested so far, including seven optical flares \citep{Graham20,Graham2023} and two gamma-ray flares \citep{Connaughton2016,Bagoly2016}.

Recently, several studies have investigated the electromagnetic emission from a variety of transients emerging from AGN disks. 
Many studies \citep{Zhu2021_Cocoon_NSMs,Zhu2021_Neutrino,Perna2021_GRBs,Yuan2021,Wang2022_Afterglow,Lazzati2022,Ray2023} 
focused on the radiation from gamma-ray bursts, while
\citet{Perna2021_AICs} and 
\citet{Zhu2021_WD_AIC} discussed the electromagnetic signatures expected from accretion-induced collapse of neutron stars and white dwarfs, respectively. 
\citet{Yang2021_TDE}, \citet{Xin2023}, and \citet{Prasad2023} studied the properties of tidal disruption of stars by BHs, 
while
\citet{Grishin2021} investigated supernova explosions, 
and \citet{Bartos17} and \citet{Stone17} estimated the electromagnetic emission produced by thermal radiation and/or outflow from circum-binary disks in AGN disks.

Several recent studies have also investigated whether transients from BHs mergingin AGN disks could explain the optical flare, ZTF19abanrhr~\citep{Graham20}, associated with the BH merger GW190521 detected in GWs.
\citet{McKernan2019_EM} discussed emission from shocks caused by 
collision between gas bound to the merger remnant and unbound gas after recoil kicks due to anisotropic radiation of GWs.
\citet{Graham20} assessed the net luminosity and timescales for gas accretion induced by 
recoil kicks. 
\citet{deMink2017} considered flares emerging from shocks in a circum-BH disk due to recoil kicks. 
\citet{Kimura2021_BubblesBHMs}, 
\citet{Wang2021_TZW,Wang2021b}, and \citet{Chen2023}, respectively, considered thermal and non-thermal emission from bubbles and bubble evolution around BHs formed by strong outflows 
considering continuous and episodic 
super-Eddington accretion. 
\citet{Wang2021_TZW} further considered emission from shocks emerging due to interactions of Blandford-Znajek jets \citep{Blandford1977} launched from accreting BHs to the broad line regions, 
\citet{RodriguezRamirez2023} considered free-free and bound-free emission from gas shocked due to interaction of the jets and AGN disk gas, 
and \citet{Tagawa2023_highenergy} estimated gamma-ray, neutrino, and cosmic-ray emission from internal shocks in the jets. 
\citet{Ashton2020}, \citet{Palmese2021}, \citet{CalderonBustillo21}, and \citet{Morton2023} estimated the association significance of ZTF19abanrhr to GW190521.

In this paper, we develop an emission model based on the scenario proposed by \citet{Tagawa2022_BHFeedback,Tagawa2023} and discuss whether or not emission based on this scenario can explain some of the optical flares reported in \citet{Graham2023}. 
\citet{Tagawa2022_BHFeedback} indicated that a Blandford-Znajek jet can be produced from BHs embedded in AGN disks and investigated its influence on the AGN disk structure. Due to the high pressure of the shocks emerging around the jet, a cavity is created around the BHs. 
Just before the jet breaks out of the AGN disk, 
photons in the shocked gas begin to escape. These photons can be observed as breakout emission \citep[e.g.][]{Nakar2010}, whose properties have been  investigated in \citet{Tagawa2023_solitary}.  
The BHs can maintain the jets even after they  break out from the AGN disk, as long as there is leftover circum-BH disk gas. Once this is depleted, the BHs can no longer power the jets.
This is then followed by an inactive phase which lasts until gas is replenished onto the BH, and the jet is launched again, with the cycle hence repeating.

\begin{figure*}
\begin{center}
\includegraphics[width=165mm]{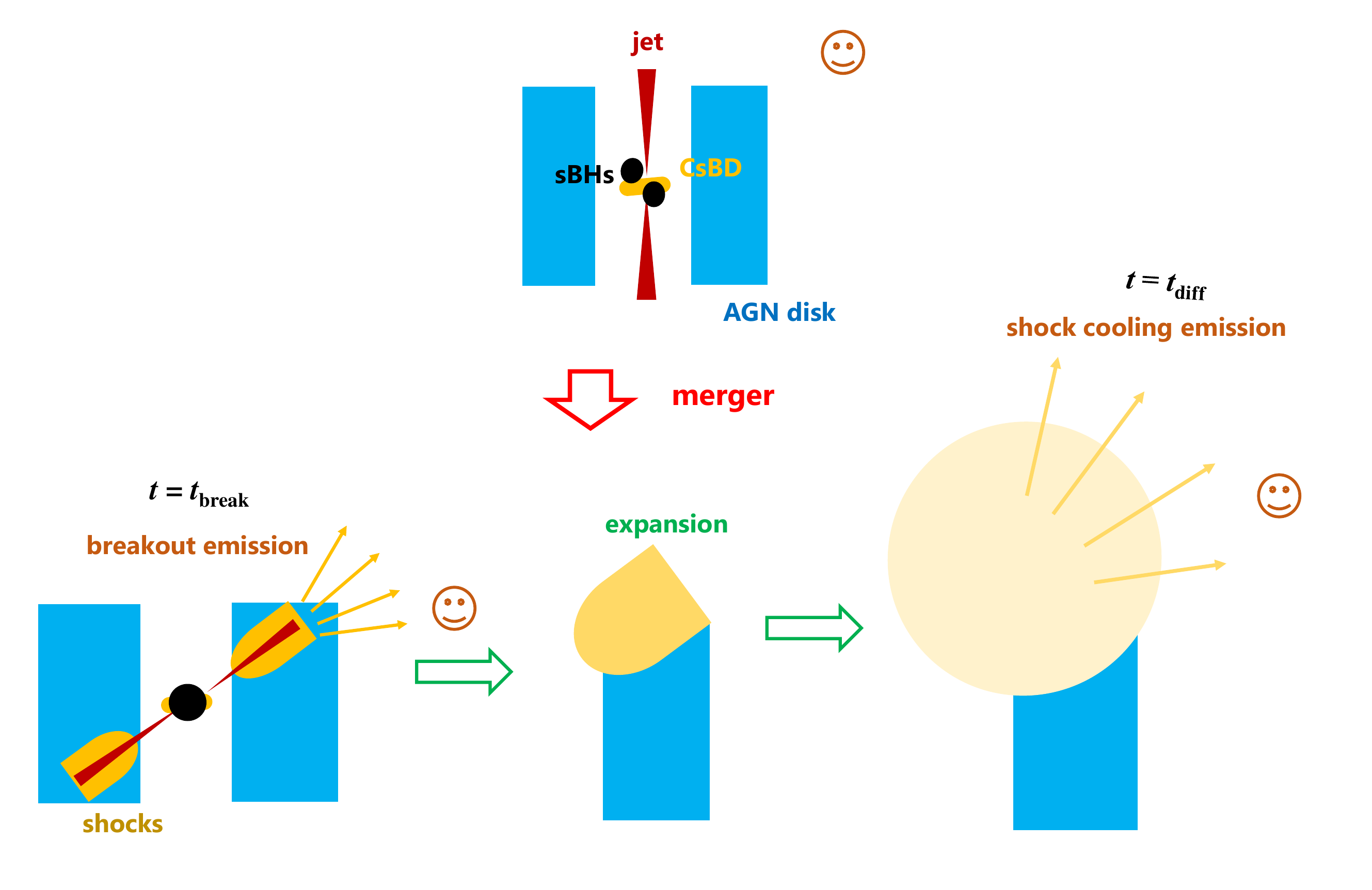}
\caption{
A schematic picture of breakout and cooling envelope emission from shocks produced by the interaction of AGN gas and a jet launched from a BH accreting gas in an AGN disk. 
Since the jet is reoriented at the time of the 
merger of two BHs ($\S~\ref{sec:model}$), electromagnetic emission is also produced after the merger. 
}
\label{fig:schematic}
\end{center}
\end{figure*}

In the case that BHs merge in the cavity while accreting (upper panel of Fig.~\ref{fig:schematic}), \citet{Tagawa2023} (hereafter Paper~I) predicted that 
electromagnetic emission is often produced associated with the BH merger. 
This is because the jet direction can be reoriented following a merger ($\S~\ref{sec:model}$), and strong shocks and emission are produced soon after the jet reorientation. 
Paper~I investigated the properties of breakout emission from a jet head associated with BH mergers, 
and found that this model can explain various properties, 
including the luminosity, delay time, duration, and color of the electromagnetic transients ZTF19abanrhr, GW150914-GBM, and LVT151012, 
as well as why the transients began brightening only after the merger. 
\citet{Zhu2023_neutrino} and 
\citet{Zhou2023} 
also estimated the neutrino emission from the breakout of the jets produced associated with BH mergers. 
In this paper, we additionally consider the shock cooling emission, which is produced in a subsequent adiabatic expansion phase of the shocked gas (e.g. \citealt[][]{Arnett1980,Morag23} for supernovae and \citealt[][]{Nakar2017,Gottliev2018,Humidani2023} for gamma-ray bursts), in addition to the breakout emission, which is produced in an early phase of the shocked gas (see Fig.~\ref{fig:schematic} for a schematic picture). 
We present the properties of the emission in both phases, and we then apply this model to 
the flares reported in \citet{Graham2023}. 
We further discuss how to test 
the proposed emission model  
in future observations.

The rest of this paper is organized as follows. 
In \S~\ref{sec:method}, 
we describe a model for producing electromagnetic flares associated with GW emission 
and a way to constrain physical properties of the flares from the observations using this model. 
We present our results in \S~\ref{sec:results}, 
discuss how to test the model in \S~\ref{sec:discussion}, and 
summarize our conclusions in \S~\ref{sec:conclusions}.

\section{Method}

\label{sec:method}

First we describe the model itself. 
We then specialize to discuss how to derive the model parameters from the observed properties of the flares, that is the delay time ($t_{\rm delay}$), the duration ($t_{\rm duration}$), the luminosity of the flare ($L_{\rm obs}$), the merger remnant mass ($m_{\rm BH}$), the 
SMBH 
mass ($M_{\rm SMBH}$), and the AGN luminosity ($L_{\rm AGN}$). 
In our analysis,
we assume that the shocks produced by collisions between the jet and the AGN gas are characterized by non-relativistic regimes. 
This is because  
in both the breakout and the shock cooling emission, flares with the delay time and the duration of $\gtrsim 10~{\rm day}$ (Table~\ref{table:properties_events}) are usually characterized by this
regime (Paper~I, Tables~\ref{table:properties_cooling} and  \ref{table:properties_model}). 
As possible processes for explaining the properties of the optical flares, 
we consider the breakout emission from the jet head (the breakout emission scenario) and shock cooling emission from the cocoon (the shock cooling emission scenario). 
Note that we use the Shakura-Sunyaev model \citep{Shakura73} for the accretion disk in the shock cooling emission scenario, 
and 
the Thompson disk model \citep{Thompson05} in the breakout emission scenario. 
This is because the position of the BH from the central supermassive BH (SMBH) is sub-parsec for the former (Table~\ref{table:properties_cooling}) and a few parsec  for the latter case (Table~\ref{table:properties_model}), and the mechanisms of angular momentum transfer and the disk properties are likely different in the two regions,
with the Thompson disk model better suited than the Shakura-Sunyaev one to describe the outer disk.

\subsection{Shock formation}

\label{sec:model}

Shocked gas is responsible for both the breakout and the cooling 
emission, and hence
we begin by describing the
process of shock formation.

We first describe the accretion rate onto BHs in an AGN disk, which 
 is evaluated by the Bondi-Hoyle-Lyttleton rate. 
For a BH embedded in a cool AGN disk, the Bondi-Hoyle-Lyttleton radius ($r_{\rm BHL}$) is large, and usually exceeds the scale height of the AGN disk ($H_{\rm AGN}$) and the Hill radius ($r_{\rm Hill}$). Accounting for the geometrical limitation of the capture regions by the shear motion and the vertical height of the AGN disk, 
the capture rate of gas by the BH is given by
\begin{align}
\label{eq:md_bondi}
{\dot m}_{\rm BHL}
=& f_{\rm c} r_{\rm w} r_{\rm h} \rho_{\rm AGN} (c_{\rm s,AGN}^2+v_{\rm BH}^2+v_{\rm sh}^2)^{1/2}\nonumber\\
\simeq & ~3\times 10^{-4} \,\Msun/{\rm yr}~
\left(\frac{f_{\rm c}}{10}\right)
\left(\frac{H_{\rm AGN}}{0.003\,{\rm pc}}\right)
\left(\frac{R_{\rm BH}}{1~{\rm pc}}\right)^{1/2}\nonumber\\
&
\left(\frac{\rho_{\rm AGN}}{4\times 10^{-17}\,{\rm g/cm^3}}\right)
\left(\frac{m_{\rm BH}}{10~\Msun}\right)^{2/3}
\left(\frac{M_{\rm SMBH}}{10^6\,\Msun}\right)^{-1/6},
\end{align}
\citep[e.g.][]{Tanigawa2016,Choksi2023},
where 
$\rho_{\rm AGN}$ is the gas density 
and $c_{\rm s,AGN}$ is the sound speed 
of the AGN disk at the position of the BH ($R=R_{\rm BH}$, where $R$ is the distance from the SMBH), 
$v_{\rm BH}$ is the velocity of the BH with respect to the local motion of the AGN disk, 
$v_{\rm sh}=r_{\rm w}\Omega$ is the shear velocity 
at the capture radius 
$r_{\rm w}={\rm min}(r_{\rm BHL},\,r_{\rm Hill})$, 
$\Omega=(GM_{\rm SMBH}/R_{\rm BH}^3)^{1/2}$ is the angular velocity of the BH, 
$r_{\rm h}={\rm min}(r_{\rm w},\,H_{\rm AGN})$ is the capture height, 
$G$ is the gravitational constant, 
and $f_{\rm c} \sim 10$ is a normalization constant \citep{Tanigawa2002}. 
In the second line of Eq.~\eqref{eq:md_bondi}, 
we assume 
$v_{\rm BH}<c_{\rm s,AGN}$ and $v_{\rm BH}<v_{\rm sh}$. 
Note that $r_{\rm w}(c_{\rm s,AGN}^2+v_{\rm sh}^2)^{1/2}\approx r_{\rm Hill}H_{\rm AGN} \Omega$ is used to derive the right hand side, which is approximately satisfied regardless of whether $c_{\rm s,AGN}$ is larger or smaller than $v_{\rm sh}$.
By considering the reduction or enhancement with respect to the Bondi-Hoyle-Lyttleton rate, 
we parameterize the fraction of the accretion rate onto the BH (${\dot m}$) over the Bondi-Hoyle-Lyttleton rate (${\dot m}_{\rm BHL}$) by $f_{\rm acc}={\dot m}/{\dot m}_{\rm BHL}$ as in Paper~I. 
For example, low $f_{\rm acc}$ may be predicted due to winds from 
an accretion disk with a super-Eddington rate. 
On the other hand, 
recent simulations suggest that the conversion to wind 
is moderate \citep{Kitaki2021} for accretion flows in which the circularization 
radius (where gas is circularized after being captured by a BH) is much larger than the trapping radius (within which photons are advected to a BH without escaping), as is the case for BHs embedded in an AGN disk.
In addition, the accretion rate onto a BH in a cavity during the active phases is estimated to be lower by a factor of a few compared to that without a cavity \citep{Tagawa2022_BHFeedback}.

From rapidly accreting and spinning BHs in an AGN disk, 
a Blandford-Znajek jet is expected to be launched, as outlined in Appendix~A.1 of \citet{Tagawa2022_BHFeedback}. 
The jet kinetic 
luminosity 
($L_{\rm j}$) is proportional to the mass accretion rate onto the BH (${\dot m}$) as 
\begin{align}
\label{eq:lj_macc}
L_{\rm j}=\eta_{\rm j}{\dot m} c^2,
\end{align}
where 
$c$ is the speed of light, and $\eta_{\rm j}$ is the jet conversion efficiency, which is approximated by $\eta_{\rm j}\sim a_{\rm BH}^2$ for a magnetically
dominated jet \citep{Tchekhovskoy2010,Narayan2021}, 
$a_{\rm BH}$ is the dimensionless spin of the BH, 
and $a_{\rm BH}\sim 0.7$ for the merger remnants \citep{Abbott21_GWTC3}.

At a BH merger ($t=0$, where $t$ is the time from the merger), 
the jet direction is reoriented and can collide with unshocked AGN gas in the following ways. 
Once two BHs merge,
the BH spin direction is reoriented if the angular momentum direction of the merging binary is misaligned with respect to the spin directions of the merging BHs. This is expected for mergers in an AGN disk due to frequent binary-single interactions \citep{Tagawa20b_spin} and/or inhomogeneity of AGN disks. 
Since the jet is injected in the direction of the BH spin, 
if the jet is not aligned with the circum-BH disk due to a strong jet power \citep{Polko2017}, 
the jet propagates in the direction of the BH spin, and can collide with AGN gas. 
Even if the jet aligns with the angular momentum direction of the circum-BH disk due to magnetic interactions on average, the jet can precess by interacting with magnetic fields 
while the angular momentum directions of the BH spin and the circum-BH disk are still misaligned with one other \citep{Liska2018,Liska2021}. Due to the precession, the jet can collide with unshocked gas after merger 
during the first precession cycle (after that, the opening angle of the cavity becomes wider than that of the precession). 
The other possibility is that once BHs merge, 
a merger remnant BH receives a recoil kick in an almost random direction due to anisotropic radiation of GWs. Then shocks form in a circum-BH disk within $\lesssim 10^{13}~{\rm cm}$ \citep{Tagawa2023}, and shocked gas accretes onto the remnant with the angular momentum direction being modified as a result of shocks \citep{Rossi2010}. 
Due to magnetic interactions \citep{Liska2021},
the jet is then aligned with the angular momentum direction of the circum-BH disk, which is in turn misaligned with respect to the jet direction before the merger,  
and the jet can therefore collide with AGN gas.

Once the jet collides with unshocked AGN gas around a BH, two shocks form: a forward shock propagating in the AGN disk and a reverse shock in the jet. The region sandwiched by the two shocks is called the jet head. 
In the jet collimated regime considered in our work, 
the dimensionless velocity of shocked gas in the jet head at the shock breakout is estimated as \citep{Bromberg2011}
\begin{align}
\label{eq:beta_h}
\beta_{\rm h}\sim 
\left(\frac{L_{\rm j}}{\rho_{\rm AGN} t_{\rm break}^2 \theta_0^4 c^5}\right)^{1/5},
\end{align}
where $\theta_0$ is the opening angle of the injected jet, and 
$t_{\rm break}$ is 
the delay time between the production of the jet and 
the shock breakout,  
and is roughly given by 
\begin{align}
\label{eq:t_break} 
t_{\rm break}&\sim \frac{3}{5} \frac{H_{\rm AGN}}{\beta_{\rm FS}c}f_{\rm corr}
\nonumber\\
&\sim1\,{\rm yr}
\left(\frac{H_{\rm AGN}}{5\times10^{16}\,\mathrm{cm}}\right)
\left(\frac{\beta_{\rm FS}}{0.1}\right)^{-1}
\left(\frac{f_{\rm corr}}{3}\right). 
\end{align}
Here $\beta_{\rm FS}\sim (7/6)\beta_{\rm h}$ is the dimensionless velocity of the forward shock of the jet head, 
and $f_{\rm corr}$ is the correction factor for the delay time due to the inclination of the jet and the geometry of the cavity. 
Ignoring 
geometrical corrections due to
the existence of the cavity, 
we assume $f_{\rm corr}={\rm min}[1/{\rm cos}i,1/\theta_0{\rm sin}i]$, 
considering 
both the cases in which
the shocked gas (cocoon) breaking out of the AGN disk is from its head and from its sides, where $i$ is the angle between the jet and the angular momentum direction of the AGN disk. 
With this prescription, $f_{\rm corr}$ ranges between $\sim 1$--$1/\theta_0$. 

After the shock breakout, 
radiation is produced. 
Early emission is characterized by the breakout emission as described in \S~\ref{sec:breakout} and Paper~I, 
while later emission is characterized by the shock cooling radiation described in \S~\ref{sec:cooling}. 
Since breakout emission is associated with the shock propagation, both non-thermal and thermal emissions are expected. 
On the other hand, we only consider thermal emission from shock cooling, and neglect any non-thermal emission from expanding ejecta. The latter can also be bright if additional strong shocks form when the ejecta collide with the interstellar medium. 
We use the non-thermal component of the breakout emission and the thermal shock cooling emission to model
the flares reported by \citet{Graham2023}, since these are bright in optical bands. 
Note that the peak frequency of the thermal breakout emission from the jet head falls above the X-ray bands, 
which is constrained by the duration of the flare (e.g. Eqs.~\ref{eq:t_bo} and \ref{eq:tem_bb0} below), and hence it cannot reproduce the optical flares.

\subsection{Shock cooling emission}
\label{sec:cooling}

\subsubsection{Physical model}

In the shock cooling emission, photons diffuse and are released from deep inside the shocked gas, 
as observed in supernovae \citep{Arnett1980}. 
To present the properties of this emission, we first describe the evolution of the shocked gas.

At the breakout of the shocked gas (cocoon) at $t=t_{\rm break}$, 
the thermal  
energy of the cocoon is
roughly given by 
\begin{align}
\label{eq:e_bo} 
e_{\rm BO}&
=\frac{1}{2}m_{\rm BO}\beta_{\rm c}^2c^2, 
\end{align}
where $\beta_{\rm c}\simeq \beta_{\rm h} \theta_0$ is the dimensionless velocity of the cocoon \citep{Bromberg2011}, and $m_{\rm BO}$ is the mass of the cocoon at the breakout and is roughly given by  
\begin{align}
\label{eq:m_bo} 
m_{\rm BO}&
\simeq 2\pi H_{\rm AGN}^3 f_{\rm corr}^3 \theta_0^2 \rho_{\rm AGN}, 
\end{align}
considering the cylindrical shape of the cocoon with the height of $H_{\rm AGN}f_{\rm corr}$ and the radius of $H_{\rm AGN}f_{\rm corr}\theta_0$. 
After the shock  
passage (even before the shock breakout), 
the internal energy of the shocked gas is converted to  kinetic energy due to the expansion 
caused
by the radiation pressure. 
Afterwards, the shocked ejecta expands nearly spherically with velocity $v_{\rm ej}\sim \beta_{\rm c}c$. 
As the ejecta expands to size $R$, 
the optical depth of the spherically expanding ejecta declines as \citep[e.g.][]{Sapir2017,Nakar2017}
\begin{align}
\label{eq:tau} 
\tau_{\rm ej}
\simeq \frac{\kappa_{\rm ej} m_{\rm BO}}{4\pi R^2}, 
\end{align}
where $\kappa_{\rm ej}$ is the ejecta's opacity. 
We adopt the Thomson scattering opacity of $\kappa_{\rm ej}\sim0.4\,{\rm g/cm^2}$ assuming ionized gas. 
If the ejecta is initially optically thick, as assumed in our fiducial model, 
photons deep inside the ejecta can be diffused out once the optical depth is reduced to 
\begin{align}
\label{eq:tau_diff} 
\tau_{\rm ej}
=c/v_{\rm ej}
\end{align} at 
the time 
\begin{align}
\label{eq:t_diff} 
t_{\rm diff}
=\left(\frac{\kappa_{\rm ej} m_{\rm BO}}{4\pi c v_{\rm ej}}\right)^{1/2}, 
\end{align}
when 
the corresponding radius is $R_{\rm diff}=t_{\rm diff}v_{\rm ej}$ 
and the density 
\begin{align}
\label{eq:rho_diff} 
\rho_{\rm diff}=\frac{c}{\kappa_{\rm ej} R_{\rm diff}v_{\rm ej}}
=\frac{c}{\kappa_{\rm ej} t_{\rm diff}v_{\rm ej}^2}.
\end{align}

Due to the adiabatic expansion, 
the thermal energy at the diffusion radius
is reduced by a factor of $\sim (R_{\rm BO}^3/R_{\rm diff}^3)^{\gamma-1}$ where $\gamma=4/3$ is the adiabatic index for the radiation pressure dominated gas, 
$R_{\rm BO}=(V_{\rm BO}/4\pi)^{1/3}$ is the typical size, and $V_{\rm BO}$ is the volume of the cocoon at the shock breakout. 
Hence, 
the luminosity at $\tau_{\rm ej}=c/v_{\rm ej}$ is 
related to $e_{\rm BO}$ as 
\begin{align}
\label{eq:l_obs} 
L_{\rm SC}=\frac{e_{\rm BO}}{t_{\rm diff}}\frac{R_{\rm BO}}{R_{\rm diff}}. 
\end{align} 
Using $R_{\rm BO}$ in Eq.~\eqref{eq:l_obs}, 
the unperturbed AGN density at the position of the BH ($R=R_{\rm BH}$) 
can be estimated via 
\begin{align}
\label{eq:rho_zero} 
\rho_{\rm AGN}\sim 
\rho_{\rm diff}\left(\frac{R_{\rm diff}}{R_{\rm BO}}\right)^3. 
\end{align} 
Furthermore, from the AGN density, 
the inflow rate of the AGN disk at $R=R_{\rm BH}$ can be calculated via
\begin{align}
\label{eq:m_inflow_alpha}
{\dot M}_{\rm inflow}=4\pi H_{\rm AGN}^3 \rho_{\rm AGN}\Omega \alpha 
\end{align}
assuming an alpha viscosity with  parameter $\alpha$, and
where $\Omega$ is the orbital angular velocity of the BH around the SMBH. 
Here, the AGN luminosity ($L_{\rm AGN}$) is 
related to the inflow rate as 
\begin{align}
\label{eq:l_md}
\frac{L_{\rm AGN}}{\eta_{\rm rad}c^2}={\dot M}_{\rm SMBH}=f_{\rm cons} {\dot M}_{\rm inflow}
\end{align} 
where ${\dot M}_{\rm SMBH}$ is the accretion rate onto the SMBH, 
$\eta_{\rm rad}$ is the radiation efficiency, 
and $f_{\rm cons}\equiv {\dot M}_{\rm SMBH}/{\dot M}_{\rm inflow}\leq1$ is the fraction of the inflow rate at $R=R_{\rm BH}$ over the accretion rate onto the SMBH.

The radiation temperature  is determined as follows. 
At $\tau_{\rm ej}=c/v_{\rm ej}$, 
the energy density of the radiation within the ejecta shell is 
\begin{align}
\label{eq:u_r} 
u_{\gamma}=\frac{L_{\rm SC}\tau_{\rm ej}}{4\pi R_{\rm diff}^2 c}. 
\end{align} 
The blackbody temperature of the radiation is then given by 
\begin{align}
\label{eq:t_bb} 
T_{\rm BB}=(u_\gamma/a)^{1/4}, 
\end{align} 
where $a$ is the radiation constant. 
Note that the radiation pressure dominates the gas pressure at $t\lesssim t_{\rm diff}$ for the models considered in this paper. 
Since the flares in \citet{Graham2023} were found by the Zwicky Transient Facility (ZTF), 
and the observed optical luminosity ($L_{\rm obs}$)
can be roughly estimated 
as 
the total observed energy of the flare 
in the optical band 
divided by the duration (provided in Table~3 of \citealt{Graham2023}), 
we pose that $L_{\rm obs}$ 
needs to match 
the luminosity in the ZTF bands as 
\begin{align}
\label{eq:lobs_lsc} 
L_{\rm obs}=L_{\rm SC} \left. \int^{\nu_{\rm up}}_{\nu_{\rm down}}\frac{2h\nu^3}{c^2}\frac{d\nu}{e^{h\nu/k_{\rm B} T_{\rm BB}}-1}\middle /\sigma T_{\rm BB}^4, \right.
\end{align} 
where $\nu_{\rm up}=c/(400~{\rm nm})$ and $\nu_{\rm down}=c/(700~{\rm nm})$ are the upper and lower limits for the frequencies observed by the $r$ and $g$ bands of the ZTF. 
Note that the luminosities in 
the $r$ and $g$ bands of the ZTF 
are not directly derived from Eq.~\eqref{eq:lobs_lsc}. 
To calculate them (panels~c and d of Fig.~\ref{fig:lc}), we adjust the limits of the integral in this equation to cover the frequency range of the relevant band.

\subsubsection{Derivation of model parameters}

Using Eqs.~\eqref{eq:md_bondi}--\eqref{eq:lobs_lsc}, 
we can then determine the 17 quantities $\beta_{\rm c}$, $\rho_{\rm AGN}$, $R_{\rm BH}$, $H_{\rm AGN}$, $L_j$, ${\dot m}_{\rm BHL}$, $t_{\rm break}$, $e_{\rm BO}$, $m_{\rm BO}$, $\tau_{\rm ej}$, $\rho_{\rm diff}$, $R_{\rm BO}$, $R_{\rm diff}$, 
$u_{\gamma}$, $T_{\rm BB}$, $L_{\rm SC}$, 
and ${\dot M}_{\rm inflow}$ given 
the 
input 
parameters, 
$\theta_0$, $f_{\rm acc}$, $a_{\rm BH}$, $\alpha$, $\eta_{\rm rad}$, $f_{\rm corr}$, and $f_{\rm cons}$, and the observed properties, $L_{\rm obs}$ $L_{\rm AGN}$, and $t_{\rm diff}$. For example, 
Combining Eqs.~\eqref{eq:md_bondi}--\eqref{eq:l_md}, 
the velocity of the cocoon is calculated as 
\begin{align}
\label{eq:beta_cooling}
\beta_{\rm c}=
\left[\frac{32 \pi^4 f_{\rm jet/BHL}^3
f_{\rm cons} A c^{10} G^2 
m_{\rm BH}^{2}t_{\rm diff}^2 \alpha \eta_{\rm rad}
}
{9
L_{\rm AGN} L_{\rm SC}^3 \kappa_{\rm ej}^4 f_{\rm corr}^{6}}\right]^{\delta},
\end{align}
where 
$f_{\rm jet/BHL} \equiv f_{\rm c} f_{\rm acc} \eta_{\rm j}$ is a 
parameter related to the jet power, and 
$A$ and $\delta$ are  variables depending on the velocity of the cocoon. 
For $\beta_{\rm c}/\theta_0<1$, 
$A=(35/18)^6 \theta_0^{-3}$
and $\delta=1/2$, 
and otherwise (as long as the jet is in the collimated regime) $A=\theta_0$ and $\delta=1/8$. 
To determine $\beta_{\rm c}$ using Eq.~\eqref{eq:beta_cooling}, 
$L_{\rm SC}$ needs to be derived using Eqs.~\eqref{eq:u_r}--\eqref{eq:lobs_lsc}. 
To consistently solve these equations, 
we calculate $\beta_{\rm c}$ in an iterative way using Newton's method. 
Then, by incorporating $\beta_{\rm c}$ into 
Eqs.\eqref{eq:md_bondi}--\eqref{eq:lobs_lsc}, 
the other parameters can also be determined.

We further adjust $f_{\rm jet/BHL}$ 
so that the scale height of the AGN disk is equal to 
the height expected for the Shakura-Sunyaev disk, given 
$\alpha$. 
Such disk structure is expected to be realized in regimes where the Toomre parameter (Eq.~\ref{eq:rho_r}) becomes greater than $1$.

\subsection{Breakout emission}
\label{sec:breakout}

\subsubsection{Physical model}

Long before photons deep inside the ejecta escape ($t\sim t_{\rm break}+t_{\rm diff}$), 
at around the time that the shock arrives to the surface of the ejecta ($t\sim t_{\rm break}$), bright breakout emission is released. 
Since breakout emission of the jet head is associated with the shock propagation, both non-thermal and thermal emission are expected. 
Such non-thermal emission can explain the optical flares found by \citet{Graham20}, as investigated in Paper~I. 
In the following we
describe how to reconstruct the model properties for the breakout emission.

Photons inside the shock start to diffuse out from the AGN disk and the breakout emission begins to be released when the photon diffusion time from the shock front, $t_{\rm diff,sh}\sim~d_{\rm edge}^2\kappa_{\rm sh}\rho_{\rm AGN}/c$, becomes equal to the dynamical timescale of the shock, $t_{\rm dyn,edge} \sim~d_{\rm edge}/\beta_{\rm FS}c$,
where $d_{\rm edge}$ is the thickness of the AGN disk above the shock, and $\kappa_{\rm sh}$ is opacity of the shocked gas. 
By equating these timescales, the thickness at the breakout 
is given by 
$d_{\rm edge,BO}\sim~1/(\beta_{\rm FS}\kappa_{\rm sh}\rho_{\rm AGN})$, and 
the duration of the emission from a breakout shell is 
\begin{align}
\label{eq:t_bo}
t_{\rm diff,BO}
&\sim 1/(\beta_{\rm FS}^2 c\kappa_{\rm sh}\rho_{\rm AGN})
\nonumber\\
&\sim3\,{\rm yr}
\left(\frac{\beta_{\rm FS}}{0.1}\right)^{-2}
\left(\frac{\rho_{\rm AGN}}{1\times10^{-16}\,{\rm g~cm^{-3}}}\right)^{-1},  
\end{align}
where we adopt 
$\kappa_{\rm sh}\sim0.4\,{\rm g/cm^2}$ 
considering the ionization of gas by photons released from the shocks \citep{Nakar2010}.

If we assume that the AGN disk is gravitationally unstable at the position of the BH, which is expected for the delay time and the duration of the flares (Paper~I), 
the density of the AGN disk is related to the position of the BH through \citep[e.g.][]{Thompson05}
\begin{align}
\label{eq:rho_r}
\frac{\Omega^2}{\sqrt{2}\pi G \rho_{\rm AGN}}=Q\simeq 1, 
\end{align}
where $Q$ is the Toomre parameter. 
Note that 
AGN disks at the BH positions are predicted to be Toomre unstable \citep[Table~\ref{table:properties_cooling}, e.g.][]{Haiman2009}.

\subsubsection{Derivation of model parameters}

Using Eqs.\eqref{eq:md_bondi}--\eqref{eq:t_break} 
and 
\eqref{eq:t_bo}--\eqref{eq:rho_r}, 
we can determine the 6 variables $\beta_{\rm h}$, $\rho_{\rm AGN}$, $R_{\rm BH}$, $H_{\rm AGN}$, $L_j$, and ${\dot m}_{\rm BHL}$, given $\theta_0$, $f_{\rm acc}$, and $\eta_{\rm j}$. 
For example, the velocity of the jet head is calculated by 
\begin{align}
\label{eq:beta}
\beta_{\rm h}=\left[\frac{5 (7/6)^{4/3} f_{\rm jet/BHL} m_{\rm BH}^{2/3}G^{1/2}\left(\frac{c \kappa_{\rm sh}t_{\rm diff,BO}}{\sqrt{2}\pi}\right)^{1/6}}{3^{5/3}f_{\rm corr}t_{\rm break}\theta_0^4 c^2}\right]^{3/11}.
\end{align}
Then, by incorporating $\beta_{\rm h}$ into 
Eqs.\eqref{eq:md_bondi}--\eqref{eq:t_break} 
and 
\eqref{eq:t_bo}--\eqref{eq:rho_r}, 
the other parameters can be also determined. 

In the case of breakout emission, 
$R_{\rm BH}$ is found to be large (\S~\ref{sec:results_breakout}). 
At such large scales, efficient transfer of angular momentum of the AGN gas is 
required for SMBH accretion 
\citep{Thompson05}. 
Following \citet{Thompson05}, 
we assume that the inflow rate of the AGN disk is parameterized by 
\begin{align}
\label{eq:m_inflow}
{\dot M}_{\rm inflow}=4\pi R_{\rm BH}H_{\rm AGN}^2 \rho_{\rm AGN}\Omega m .
\end{align}

\subsection{Evolution}

In this section, we describe the evolution of the luminosity and the temperature for thermal emission from the cocoon in the breakout and shock cooling emission phases. 
Referring to previous studies \citep{Arnett1980,Nakar2010,Piro2021,Morag23}, 
we assume that the luminosity evolves as 
\begin{eqnarray}
\label{eq:l_ev}
&L(t')
\sim
\frac{e_{\rm BO}}{t_{\rm dyn}} 
&\left\{
\begin{array}{l}
0~~~~\mathrm{for}~~t'\lesssim 0 ,\\
1~~~~\mathrm{for}~~0\lesssim t'\leq t_{\rm pl} ,\\
\left(\frac{t'}{t_{\rm pl}}\right)^{-4/3}
~~~~\mathrm{for}~~t_{\rm pl}\leq t' \lesssim t_{\rm sph} ,\\
\left(\frac{t_{\rm sph}}{t_{\rm pl}}\right)^{-4/3}
\left(\frac{t'}{t_{\rm sph}}\right)^{-\frac{2.28n-2}{3(1.19n+1)}}\\
~~~~~~~~~~~~\mathrm{for}~~t_{\rm sph}\leq t' \leq t_{\rm diff},\\
\frac{t_{\rm dyn}}{t_{\rm diff}}
\frac{R_{\rm BO}}{R_{\rm diff}}
{\rm exp}\left[-\frac{1}{2}\left(\frac{t'^2}{t_{\rm diff}^2}-1\right)\right]\\
~~~~~~~~~~~~\mathrm{for}~~t_{\rm diff}\leq t' ,
\end{array}
\right.
\end{eqnarray}
where $t'=t-t_{\rm break}$, 
$t_{\rm pl}=1/(\beta_{\rm c}^2 c \kappa_{\rm ej}\rho_{\rm AGN})$ is the duration of emission from a breakout shell in the cocoon, 
$t_{\rm sph}\simeq R_{\rm BO}/\beta_{\rm c}c$ is the transition between the planar and spherical geometries of the breakout shell, 
$t_{\rm dyn}=H_{\rm AGN}/\beta_{\rm c} c$ is the dynamical timescale,  
and $n$ is the power law slope of the vertical
AGN gas density profile 
at the height at which photons begin to break out.  
The second and third rows correspond to the luminosity in the planar phase (before the shocked gas doubles its radius by expansion) for the breakout emission, the fourth row corresponds to that in a spherical phase (after the shocked gas doubles and before the shock cooling emission phase), and the fifth row corresponds to that in a shock cooling emission phase. 

Similarly following previous studies, we assume that the temperature evolves as
\begin{eqnarray}
\label{eq:t_ev}
&T(t')
\sim
T_{\rm BB,0} 
&\left\{
\begin{array}{l}
1~~~~\mathrm{for}~~0\lesssim t'\leq t_{\rm pl} ,\\
\left(\frac{t'}{t_{\rm pl}}\right)^{-\frac{2}{3}\frac{9n+5}{17n+9}}
~~~~\mathrm{for}~~t_{\rm pl}\leq t' \lesssim t_{\rm sph} ,\\
\left(\frac{t_{\rm sph}}{t_{\rm pl}}\right)^{-\frac{2}{3}\frac{9n+5}{17n+9}}
\left(\frac{t'}{t_{\rm sph}}\right)^{n_{\rm T}}\\
~~~~~~~~~~~~\mathrm{for}~~t_{\rm sph}\leq t' \leq t_{\rm diff},\\
\frac{T_{\rm BB}}{T_{\rm BB,0}}
\left(\frac{t'^2}{t_{\rm diff}^2}-1\right)^{-1/2}
~~~~\mathrm{for}~~t_{\rm diff}\leq t' ,
\end{array}
\right.
\end{eqnarray}
where 
\begin{eqnarray}
\label{eq:tem_bb0}
T_{\rm BB,0}=(7\rho_{\rm AGN} \beta_{\rm c}^2 c^2/2a)^{1/4} 
\end{eqnarray}
is the black body temperature of the breakout shell, 
$n_{\rm T}$ is the power-law index for the temperature evolution in the spherical phase, and 
$n_{\rm T} =-(18.48n^2+20.69n+6)/[(1.19n+1)(22.32n+17)]$ for an expanding spherical ejecta \citep{Nakar2010}. 
The validity of these analytical formulae has been tested by recent numerical simulations \citep[e.g.][]{Piro2021,Morag23}.

\begin{table*}
\begin{center}
\caption{The properties of the ZTF flares suggested to be associated with the GW events. 
In each pair, 
the flare is located within the $90\%$ credible volume of 
and detected within the delay time of $200$ days
from the GW event. 
Here, 
$M_{\rm SMBH}$ is the SMBH mass, 
$m_{\rm BH}$ is the merged remnant mass, 
$t_{\rm duration}$ is the duration of the flare, 
$t_{\rm delay}$ is the delay time between the detection of the GWs to and the flare, 
$L_{\rm op}$ is the optical luminosity of the flare, 
and $L_{\rm AGN}$ is the Eddington ratio of the host AGN. 
For $\rm J053408.41+085450.6$, the SMBH mass is not constrained and a fiducial mass of $10^8~\Msun$ is used as in \citet{Graham2023}. 
}
\label{table:properties_events}
\begin{tabular}{c|c|c|c|c|c|c|c|c}
\hline 
Pair &LIGO/Virgo alart ID&Name of AGN&
${\rm log_{10}}\left(\frac{M_{\rm SMBH}}{\Msun}\right)$&$m_{\rm BH}~[\Msun]$
& $t_{\rm duration}~[\rm day]$& $t_{\rm delay}~[{\rm day}]$&$L_{\rm op}~[\rm erg/s]$&$L_{\rm AGN}/L_{\rm Edd}$\\
\hline\hline
1&$\rm GW190403\_051519$&$\rm J124942.30+344928.9$&8.6&111&45.3&89&44.9&0.06\\\hline
2&$\rm GW190514\_065416$&J124942.30+344928.9&8.6&68&45.3&48&44.9&0.06\\\hline
3&$\rm GW190521$&J124942.30+344928.9& 8.6 &164& 45.3 &41&44.9&0.06\\\hline
4&$\rm GW190403\_051519$&J183412.42+365655.3&9.1&111& 41.0&110&43.8&0.02\\\hline
5&$\rm GW190424\_180648$&J181719.94+541910.0&8.0&73&35.6 &189&44.8&0.05\\\hline
6&$\rm GW190514\_065416$&J224333.95+760619.2&8.8&68&18.3&154&45.2&0.06\\\hline
7&$\rm GW190731\_140936 $&J053408.41+085450.6 & (8.0) &70.1&27.4 &186&44.5&0.04\\\hline
8&$\rm GW190803\_022701$&J053408.41+085450.6& (8.0)&65& 27.4 &183&44.5&0.04\\\hline
9&$\rm GW190803\_022701$&J120437.98+500024.0& 8.0&65& 47.4 &191&44.7&0.02\\\hline
10&$\rm GW190909\_114149$&J120437.98+500024.0& 8.0&75& 47.4 &154&44.7&0.02\\\hline
11&$\rm GW200216\_220804$&J154342.46+461233.4& 9.3&81& 123.4 &73&44.3&0.10\\\hline
12&$\rm GW200220\_124850$&J154342.46+461233.4& 9.3&67& 123.4&69&44.3&0.10\\\hline
\end{tabular}
\end{center}
\end{table*}

Since Eqs.~\eqref{eq:l_ev} and \eqref{eq:t_ev} are formulae for an expanding ejecta with a spherically symmetric evolution, some modifications are required in the cases of emission from an expanding ejecta with a cylindrical shape like a cocoon, especially at around the transition between the planar and spherical phases. 
For simplicity, we determine $n$ and $n_{\rm T}$ so that the luminosity and the temperature smoothly evolve at $t'=t_{\rm diff}$, respectively.

\subsection{Parameters}

\label{sec:parameters}

In this section we describe the fiducial values for the model parameters, and the observed properties of the flares used in the modellings. 

As fiducial values, we set 
the opening angle of the injected jet to $\theta_0=0.2$ \citep[e.g.][]{Hada2013,Hada2018,Berger2014}, 
the radiation efficiency to $\eta_{\rm rad}=0.1$, 
the correction factor for the delay time to $f_{\rm corr}=3$ (which roughly corresponds to the median value considering the cavity with the aspect ratio of $\sim 1$ and isotropic jet directions), 
the consumption fraction of the inflow rate to $f_{\rm cons}=1$, 
the alpha-viscous parameter to $\alpha=0.1$, 
and the angular momentum transfer parameter in the outer regions of the AGN disk to $m=0.1$. 
For the computation of the shock cooling emission we 
adjust the value of 
$f_{\rm jet/BHL} \equiv f_{\rm c} f_{\rm acc} \eta_{\rm j}$ 
so that the scale height of the AGN disk is equal to the height expected for the Shakura-Sunyaev disk (\S~\ref{sec:cooling}), 
while 
for the breakout emission
we set it to 
$f_{\rm jet/BHL}=10$ 
assuming $f_{\rm c}=10$ \citep{Tanigawa2002}, $\eta_{\rm j}=0.5$, and $f_{\rm acc}=2$ considering moderate enhancement of accretion due to shocks caused by recoil kicks \citep[Appendix~B,][]{Tagawa2023}.

When modeling the flares with breakout emission, 
we assume that the flare duration ($t_{\rm duration}$) corresponds to the exponential decay time ($t_{e}$) in \citet{Graham2023}. 
To derive the delay time ($t_{\rm delay}$), we identify the day at which the flare luminosity peaks using {\it digitizer}\footnote{https://automeris.io/WebPlotDigitizer/}. 
Due to the difficulty of identifying the peak, 
$t_{\rm delay}$ contains uncertainties. 
We adopt 
$M_{\rm SMBH}$ and $m_{\rm BH}$ from Tables~3 and 4 of \citet{Graham2023}. 
The optical luminosity is derived by means of 
the total observed energy 
in the optical band 
divided by the sum of the rise time ($t_{g}$) and the decay time ($t_{e}$) presented in \citet{Graham2023}. 
We calculate the luminosity of the AGNs ($L_{\rm AGN}$) by inferring the flux of the AGNs 
at $\sim 4000${\AA} 
from Fig.~6 of \citet{Graham2023}, 
and estimating the luminosity distance from the redshift of the GW events assuming a value for the Hubble constant of $67.8~{\rm km/s}~{\rm s}^{-1}~{\rm Mpc}^{-1}$, 
for the matter density today of 0.24, and for the cosmological constant today of 0.74 \citep{Planck2016}, and adopting the bolometric correction factor of $5$ \citep{Duras2020}. 
The values for the observed quantities adopted in this paper are listed in Table~\ref{table:properties_events}. 
Note that the delay time ($t_{\rm delay}$) corresponds to $t_{\rm break}$, and the duration of a flare ($t_{\rm duration}$) corresponds to $t_{\rm diff,BO}$.

Conversely, when modeling the flares with shock cooling emission,
we assume that 
the duration of a flare ($t_{\rm duration}$) corresponds to $t_{\rm diff}$. 
The delay time between a GW event and an optical flare ($t_{\rm delay}$) 
is also on the order of $\sim t_{\rm diff}$ 
given $t_{\rm break}\ll t_{\rm diff}$, 
while we do not use the delay time ($t_{\rm delay}$) to constrain the model parameters. 
Note that $t_{\rm delay}\sim t_{\rm duration}$ expected in this scenario is roughly consistent with the properties of the observed flares  (table~\ref{table:properties_events}). 
We assume that the observed optical luminosity of the flare ($L_{\rm op}$) corresponds to 
the luminosity in the ZTF bands ($L_{\rm obs}$).

We note that in our model, the
physical properties are uniquely determined. 
This is because we fixed several input parameters as detailed above, 
and we do not directly use the observational data points for parameter fitting. 
If instead we allowed the input parameters to vary, we would introduce degeneracies among several input parameters.
The variations of the values of the fixed input parameters can affect physical properties of the model especially in the breakout-emission 
scenario, while less affect them in the shock cooling emission scenario (see \S~\ref{sec:dis_cooling} and \S~\ref{sec:dis_breakout}). 
We believe that such simple prescriptions are useful to understand typical properties expected from the model and consider possible tests of the model below.

\begin{table*}
\begin{center}
\caption{
The inferred model parameters 
for 
the shock cooling emission scenario. 
$R_{\rm BH}$ is the distance from the SMBH to the BH, 
$\beta_{\rm c}$ is the expansion velocity of the cocoon, 
$h_{\rm AGN}\equiv H_{\rm AGN}/R_{\rm BH}$ and $\rho_{\rm AGN}$ are the aspect ratio and the density of the AGN disk at the position of the BH, respectively, 
$T_{\rm BB}$ and $\lambda_{\rm peak}$ are the radiation temperature and wavelength of the shock cooling emission, respectively, 
$L_{\rm jet}$ is the jet power, 
$f_{\rm jet/BHL}$ is a parameter related to the
jet power, 
$t_{\rm break}$ is the breakout timescale of the jet, 
and $t_{\rm diff,CBO}$ is the diffusion timescale for the breakout emission from the cocoon. 
}
\label{table:properties_cooling}
\begin{tabular}{c|c|c|c|c|c|c|c|c|c|c|c}
\hline 
Pair &
$R_{\rm BH}~[{\rm pc}]$&
$R_{\rm BH}~[R_{\rm g}]$&
$\beta_{\rm c}$&
$h_{\rm AGN}$&$\rho_{\rm AGN}~[\rm g/cm^3]$&
$T_{\rm BB}~[{\rm K}]$&$\lambda_{\rm peak}~[{\rm nm}]$
&$L_{\rm jet}~[{\rm erg/s}]$
&
$f_{\rm jet/BHL}$
&$t_{\rm break}~[\rm day]$
&$t_{\rm diff,CBO}~[\rm s]$ (Eq.~\ref{eq:t_diff_cbo})
\\
\hline\hline
1 & 0.1
&$5\times 10^3$
& 0.2 & $0.004$ & 1$\times 10^{-11}$ & $1.4\times 10^4$ 
&350
& $1\times 10^{48}$
&40
&0.9
&200
\\\hline
2 & 0.1
&$5\times 10^3$
& 0.2 & 0.004 & 1$\times 10^{-11}$ & $1.4\times 10^4$ 
&350
& $1\times 10^{48}$
&50
&0.9
&200
\\\hline
3 & 0.1
&$5\times 10^3$
& 0.2 & 0.004 & 1$\times 10^{-11}$ & $1.4\times 10^4$ 
&350
& $1\times 10^{48}$ 
&30
&0.9
&300
\\\hline
4 & 0.08
&$1\times 10^3$
& 0.08
& 0.002 & 7$\times 10^{-11}$ 
& $1.4\times 10^4$ &350
& $1\times 10^{47}$ 
&2
&0.8
&200
\\\hline
5 & 0.1
&$3\times 10^4$
& 0.2
& 0.005 & 1$\times 10^{-12}$ 
& $1.8\times 10^4$
& 270
&$2\times 10^{47}$ 
&40
&2
&$3 \times10^3$
\\\hline
6 & 0.06
&$2\times 10^3$
& 0.4
& 0.003 & 9$\times 10^{-11}$ 
& $1.3\times 10^4$ &380
& $2\times 10^{50}$ 
&3000
&0.2
&7
\\\hline
7 & 0.1
&$2\times 10^4$
& 0.1 & 0.005 & 2$\times 10^{-12}$ 
& $1.9\times 10^4$ 
& 250
&$2\times 10^{47}$
&30
&1
&$2 \times10^3$
\\\hline
8 & 0.1
&$2\times 10^4$
& 0.1 & 0.005 & 2$\times 10^{-12}$
& $1.9\times 10^4$ 
&250
& $2\times 10^{47}$ 
&30
&1
&$2 \times10^3$
\\\hline
9 & 0.3
&$6\times 10^4$
& 0.1
& 0.005 & 2$\times 10^{-13}$ 
& $2.0\times 10^4$ 
&240
& $5\times 10^{46}$ 
&20
&6
&$5 \times10^4$
\\\hline
10 & 0.3
&$6\times 10^4$
& 0.1
& 0.005 & 2$\times 10^{-13}$
& $2.0\times 10^4$ 
&240
&$5\times 10^{46}$ 
&20
&6
&$5\times10^4$
\\\hline
11 & 0.09
&$9\times 10^2$
& 0.09 
& 0.003 & 2$\times 10^{-10}$ & $9.2\times 10^3$ & 520&
$8\times 10^{47}$
&5
&1
&60
\\\hline
12 & 0.09
&$9\times 10^2$
& 0.09
& 0.003 & 2$\times 10^{-10}$ & $9.2\times 10^3$ &520
& $8\times 10^{47}$
&5
&1
&50
\\\hline
\end{tabular}
\end{center}
\end{table*}

\begin{table*}
\begin{center}
\caption{
The inferred model parameters 
for the breakout emission 
scenario. 
${\dot M}_{\rm inflow}/{\dot M}_{\rm Edd}$
is the Eddington ratio for the gas inflow at the position of the BH, 
$f_{\rm op/kin}$ is the fraction of the optical luminosity of the flare over the kinetic power of the jet, 
$f_{\rm BH/NSC}$ is 
the fraction of $R_{\rm BH}$ over the size of the nuclear star cluster, and 
$h_{\rm AGN}/h_{\rm AGN,TQM}$ is the ratio of 
the aspect ratio of the AGN disk derived from our model to that derived assuming the model in \citet{Thompson05}. 
In the last column, the value of the correction factor for the delay time, at which $h_{\rm AGN}=h_{\rm AGN,TQM}$ is satisfied, is listed. 
}
\label{table:properties_model}
\hspace{-5mm}
\begin{tabular}{c|c|c|c|c|c|c|c|c|c}
\hline 
Pair &
$R_{\rm BH}~[{\rm pc}]$&$\beta_{\rm h}$&
$h_{\rm AGN}$&$\rho_{\rm AGN}~[\rm g/cm^3]$&
${\dot M}_{\rm inflow}/{\dot M}_{\rm Edd}$&$f_{\rm op/kin}$&$f_{\rm BH/NSC}$&$h_{\rm AGN}/h_{\rm AGN,TQM}$
&$f_{\rm corr}$ for $h_{\rm AGN}=h_{\rm AGN,TQM}$
\\
\hline\hline
1 & 3.3 & 0.31 & 0.004 & $2\times 10^{-16}$ & 0.06 & 0.36 & 0.38 &1.05&3.1 \\\hline
2 & 3.5 & 0.33 & 0.002 & $1\times 10^{-16}$ & 0.02 & 0.98 & 0.41 & 1.6&4.3\\\hline
3 & 4.0 & 0.41 & 0.002 & $1\times 10^{-16}$ & 0.01 & 0.73 & 0.46 &1.5&4.2 \\\hline
4 & 4.5 & 0.29 & 0.004 & $2\times 10^{-16}$ & 0.05 & 0.02 & 0.44 &1.03&3.1 \\\hline
5 & 1.6 & 0.23 & 0.015 & $4\times 10^{-16}$ & 0.9 & 0.11 & 0.23 & 1.4&4.0\\\hline
6 & 2.4 & 0.23 & 0.008 & $7\times 10^{-16}$ & 0.4 & 0.22 & 0.26 &1.6&4.4 \\\hline
7 & 1.4 & 0.23 & 0.016 & $5\times 10^{-16}$ & 1.2 & 0.05 & 0.21 &1.6&4.3 \\\hline
8 & 1.4 & 0.22 & 0.016 & $5\times 10^{-16}$ & 1.2 & 0.05 & 0.21&1.6&4.3 \\\hline
9 & 1.7 & 0.23 & 0.014 & $3\times 10^{-16}$ & 0.7 & 0.14 & 0.25 &1.3&3.7 \\\hline
10 & 1.8 & 0.25 & 0.011 & $3\times 10^{-16}$ & 0.4 & 0.16 & 0.26 &1.1&3.3 \\\hline
11 & 8.1 & 0.32 & 0.002 & $6\times 10^{-17}$ & 0.004 & 0.34 & 0.74&2.0&5.2 \\\hline
12 & 8.0 & 0.31 & 0.001 & $6\times 10^{-17}$ & 0.004 & 0.41 & 0.73 &1.9&4.9 \\\hline
\end{tabular}
\end{center}
\end{table*}

\begin{figure*}\begin{center}
\includegraphics[width=180mm]{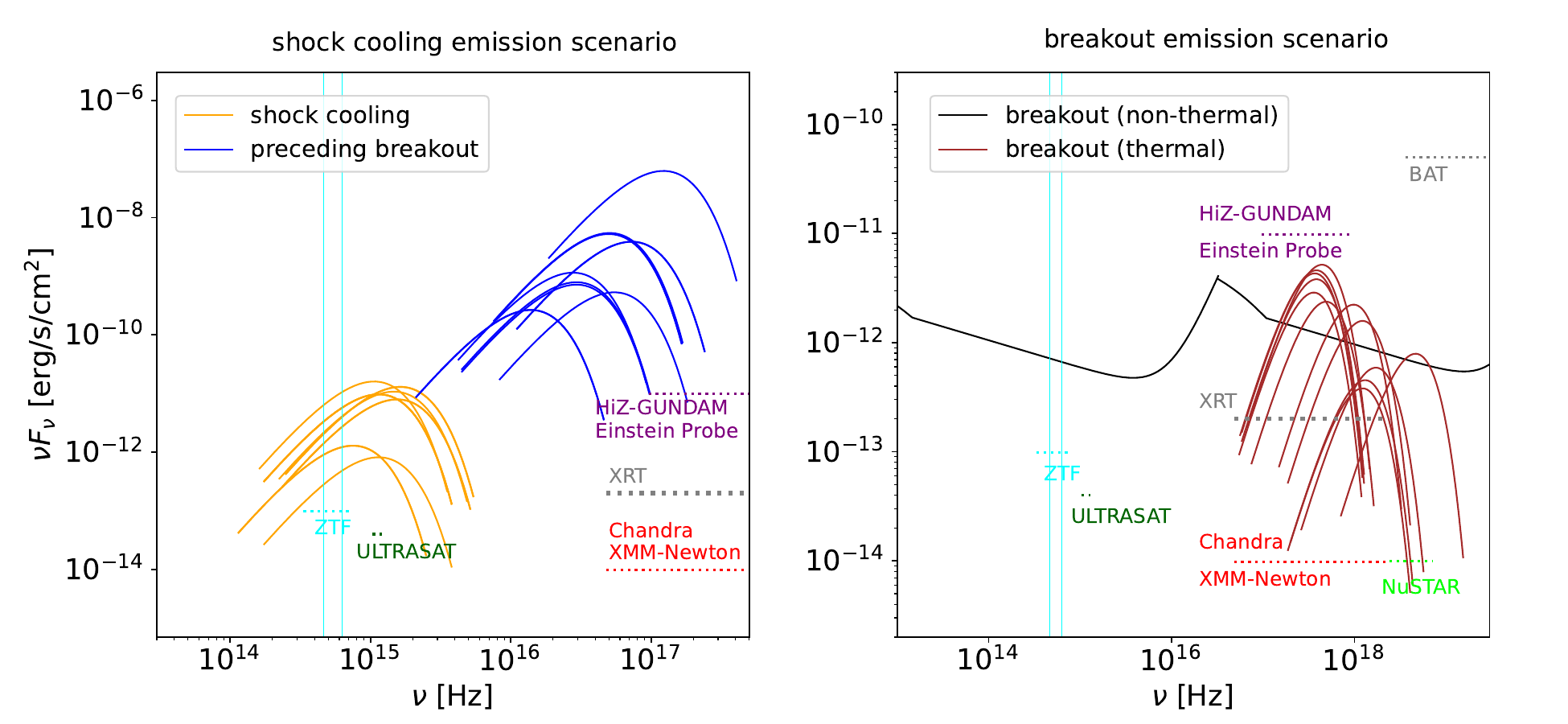}
\caption{
The SEDs for shock cooling (orange) and preceding breakout emission 
(blue) from a cocoon in the shock cooling emission 
scenario 
for pairs 1--12 (left panel), 
and those for thermal breakout emission (brown) from a jet head in the breakout emission 
scenario 
(right panel) 
assuming the luminosity distance of $d_{\rm L}=3~{\rm Gpc}$. 
The spectral energy distributions (SEDs) for the non-thermal emission is drawn for the model adopted in Fig.~5 of Paper~I as a representative example. 
The dotted blue, purple, red, green, dark green, and gray lines mark 
the sensitivity of the ZTF, HiZ-GUNDAM and Einstein Probe, Chandra and XMM-Newton, NuSTAR, ULTRASAT, 
and Swift BAT, respectively, as adopted in Paper~I. 
The vertical cyan lines present the $g$ and $r$ bands of ZTF. 
Note that the shock cooling emission is too dim ($\ll 10^{-17}~{\rm erg/s/cm^2}$) in the breakout emission scenario 
to be visible in the right panel. 
}\label{fig:sed}\end{center}\end{figure*}

\begin{figure*}\begin{center}
\includegraphics[width=180mm]{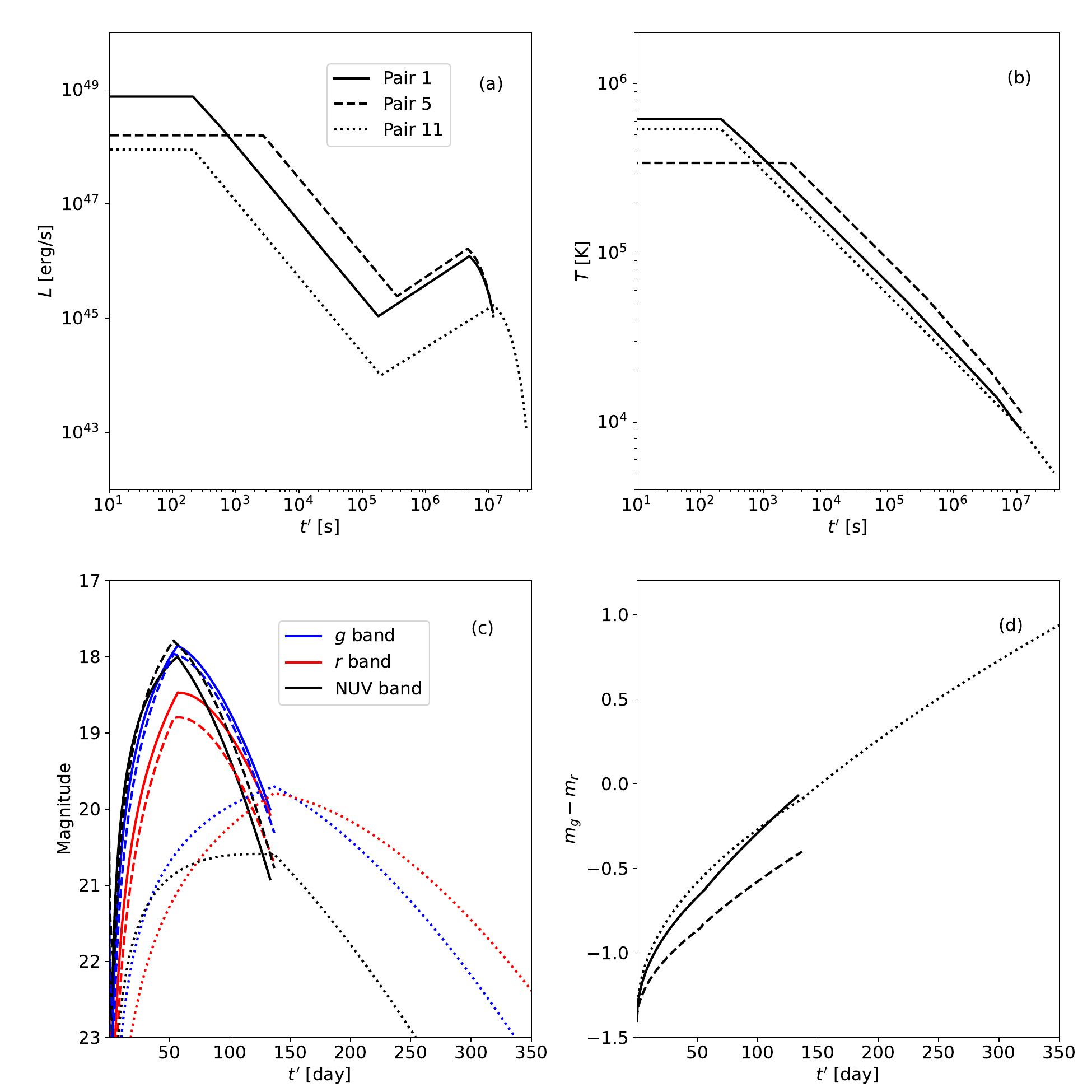}
\caption{
The evolution of the luminosity (panel~a), temperature (panel~b), 
apparent magnitude in the $g$ ($m_g$, blue) and $r$ ($m_r$, red) bands, 
AB magnitude at a near-ultraviolet band (260$~{\rm nm}$, black, panel~c), 
and the color ($m_g-m_r$, panel~d) 
for thermal emission in the shock cooling emission model for pairs~1 (solid), 5 (dashed), and 11 (dotted). 
We present 
pairs~1, 5, and 11 as a fiducial, 
a smaller $m_{\rm BH}$, 
and a longer $t_{\rm duration}$ case, respectively, 
among which evolution timescales are derived to be different. 
Lines are drawn until $\tau_{\rm ej}=1$ after which our model is invalid. 
}\label{fig:lc}\end{center}\end{figure*}

\section{Results}

\label{sec:results}

\subsection{Shock cooling emission}

\label{sec:results_cooling}

Table~\ref{table:properties_events} shows 
the list of possible pairs of associations between the GW events and the electromagnetic flares reported in \citet{Graham2023}, along with their observed properties. 
Seven electromagnetic flares and twelve pairs for the possible associations are reported. 
The number of the pairs (12) is larger than that of the flares (7), 
because some flares can be associated with more than one GW events. 
Hence, at least five pairs are false associations. 
Note that in this paper we do not analyze the two gamma-ray flares possibly associated with GW events due to their significantly different properties \citep{Connaughton2016,Bagoly2016}, while they can be explained by the thermal breakout emission from the jet head in relativistic regimes \citep{Tagawa2023}.

Table~\ref{table:properties_cooling} shows 
the distribution of the model parameters when the properties of the observed flares (Table~\ref{table:properties_events}) are modeled with
shock cooling emission.

The parameter
$f_{\rm jet/BHL}$ is widely distributed depending on the pairs. 
$f_{\rm jet/BHL}\sim1$--$10$ in pairs 4, 11, and 12 is roughly expected for the Bondi-Hoyle-Lyttleton accretion, while $f_{\rm jet/BHL}\sim10$--$60$ in pairs 1--3, 5, and 7--10 can be realized by the enhancement of accretion by shocks due to recoil kicks at merger (Paper~I). 
However, $f_{\rm jet/BHL}\sim3000$ in pair 6 is difficult to
be realized, 
since only up to $f_{\rm jet/BHL}\lesssim300$ is found to be feasible by considering the radial surface density profile of a circum-BH disk and appealing to the results of numerical simulations (see Appendix~B of Paper~I). 
On the other hand, 
even for pair~6, if we adopt $f_{\rm corr}\sim 1$, 
$f_{\rm jet/BHL}$ is reduced to $\sim 300$. 
This is presumably because both a high accretion rate and a low inclination have similar influences on the breakout velocity, and most parameters are mainly characterized by the magnitude of the breakout velocity. 
Due to the variation of $f_{\rm corr}$ (between $\sim 1$--$1/\theta_0$) 
and the degeneracy between $f_{\rm corr}$ and $f_{\rm jet/BHL}$, 
the cooling model is poorly constrained or tested by the value of $f_{\rm jet/BHL}$. 

The parameter $R_{\rm BH}$ ranges between $0.06$--$0.3$ pc, corresponding to $900$--$6\times 10^4~R_g$, where $R_{\rm g}=GM_{\rm SMBH}/c^2$ is the gravitational radius of the SMBH. 
These locations roughly correspond to migration traps, gap forming regions, and/or slow migration regions, where BH mergers are predicted to be frequent \citep{Bellovary16,Tagawa19}.

The orange lines in the left panel of Fig.~\ref{fig:sed} show the SEDs for pairs 1--12. 
The radiation temperature ranges between $T_{\rm BB}=9200$--$20000~{\rm K}$, 
corresponding to 
the peak wavelength $\lambda_{\rm peak}=ch/(3k_{\rm B}T_{\rm BB})=240$--$520~{\rm nm}$, 
where $h$ is the Planck constant and $k_{\rm B}$ is the Boltzmann constant. 
Also, the various colors observed in the optical flares (Fig.~3 of \citealt{Graham2023}) can be reproduced depending on the variation 
of 
the radiation temperature 
with changing some input parameters, such as the direction of the jet ($i$). 
For example, 
if we set $i=0^{\circ}$ 
(the jet being perpendicular to the AGN-disk plane) 
by using derived values for $\rho_{\rm AGN}$, $R_{\rm BH}$, and $H_{\rm AGN}$ and the observed parameters ($m_{\rm BH}$ and $M_{\rm SMBH}$), the radiation temperature of flares is enhanced by a factor of $\sim 1.7$. This is because for $i=0^\circ$ the shocked mass becomes low, and then photons can escape from an earlier phase when the radiation temperature is higher. 
Thus, the variety of colors can be reproduced if there are temperature variations in the thermal emission. 
For synchrotron emission 
(whose possible contribution to the flares is discussed in \S~\ref{sec:results_breakout}), 
the color is related to 
the power-law slope of the injected electrons accelerated by the first-order Fermi process ($p$) as $\nu L_\nu \propto \nu^{(-p+2)/2}$, 
and $p$ is presumably distributed in a narrow range for the same phenomena \citep{Panaitescu2001}. 
On the other hand, various values for the color can be realized if the breaks of the power laws in the SEDs coincidentally fall at around the ZTF bands (see the black line in the right panel of Fig.~\ref{fig:sed}). 
Note that the distribution of the colors is currently uncertain due to the shift in the baseline of the flux and the contribution from the background AGN emission in Fig.~3 of \citet{Graham2023}. 
If the color is actually distributed in a wide range, the flares are then easier to model by thermal shock cooling emission rather than by non-thermal breakout emission. 
In addition, the model can be also discriminated by constraining the spectral shape of the emission. This can be done by simultaneously observing the flares by the ZTF and the Ultraviolet Transient Astronomy Satellite 
(ULTRASAT, \citealt{Shvartzvald2023}, Fig.~\ref{fig:sed}) in the future.

We also predict 
breakout emission from the cocoon preceding the shock cooling emission 
(this is the emission from the cocoon, which is different from the emission from the jet head discussed in \S~\ref{sec:results_breakout}). 
The duration of the breakout emission from the cocoon, 
\begin{align}
\label{eq:t_diff_cbo} 
t_{\rm diff,CBO}=\frac{1}{\beta_{\rm c}^2 c \kappa_{\rm ej}\rho_{\rm AGN}},
\end{align}
is distributed in the range $t_{\rm diff,CBO}\sim 10$--$10^5~{\rm s}$ (the right most column of Table~\ref{table:properties_cooling}), and the temperature 
(Eq.~\ref{eq:tem_bb0}) 
is in a range $2\times 10^5$--$2\times 10^6~{\rm K}$ (blue lines in the right panel of Fig.~\ref{fig:sed}). 
Although the duration of the breakout emission is shorter than than that of the shock cooling emission, it 
can be detected by future X-ray surveys, such as HiZ-GUNDAM \citep{Yonetoku2020c} and/or the Einstein Probe \citep{Yuan2014}.

Fig.~\ref{fig:lc}~a and b show the evolution of the luminosity and the temperature of emission for pairs~1, 5, and 11. 
We chose pair~1 as a fiducial example, 
pair~5 as an event with small $m_{\rm BH}$, 
and pair~11 as the one with long $t_{\rm duration}$.  
The lines are drawn until the phase at which $\tau_{\rm ej}=1$ is satisfied, 
since our model for thermal emission is not valid when $\tau_{\rm ej}<1$.

For the smaller $m_{\rm BH}$ case (dashed lines in Fig.~\ref{fig:lc}), 
the timescale for the breakout emission of the cocoon ($t_{\rm pl}$) 
is 
longer. 
This is because $L_{\rm j}$ is lower due to the lower Bondi-Hoyle-Lyttleton rate (Eqs.~\ref{eq:md_bondi}, \ref{eq:lj_macc}), and $\beta_{\rm c}$ tends to be lower for lower $L_{\rm j}$ (Eqs.~\ref{eq:beta_h}, \ref{eq:beta_cooling}). To reproduce $t_{\rm diff}$, $\rho_{\rm AGN}$ 
needs to be 
lower for lower $\beta_{\rm c}$ (Eq.~\ref{eq:t_diff}). Due to the low $\beta_{\rm c}$ and $\rho_{\rm AGN}$ (Table~\ref{table:properties_cooling}), the breakout timescale of the cocoon ($t_{\rm pl}\propto \beta_{\rm c}^{-2} \rho_{\rm AGN}^{-1}$) becomes longer. 
In such case, 
since photons diffuse out from the shocked gas earlier, adiabatic expansion is inefficient, leading to higher temperature.

For the longer $t_{\rm duration}$ model (dotted lines in Fig.~\ref{fig:lc}), the evolution of the luminosity is slower in the shock cooling regime ($t'\gtrsim t_{\rm diff}$).

\subsection{Breakout emission}

\label{sec:results_breakout}

Table~\ref{table:properties_model} shows 
the parameters of the breakout emission model 
reproducing the properties of the observed flares. 
In this model, the  parameters are distributed in narrow ranges. 
The mergers are predicted to occur at 
$R_{\rm BH}\sim 1$--$10~{\rm pc}$, 
the shock velocity ranges within $\beta_{\rm h}\sim 0.3$--$0.4$, 
the disk aspect ratio at $R_{\rm BH}$ ranges within $h_{\rm AGN}\equiv H_{\rm AGN}/R_{\rm BH}\sim 0.001$--$0.02$, 
the fraction of the optical luminosity to the kinematic power of the shock is around $L_{\rm op}/L_{\rm j}\sim 0.02$--$1$, 
the fraction of the distance from the SMBH to the size of the nuclear star cluster ($R_{\rm NSC}$) ranges within $R_{\rm BH}/R_{\rm NSC}\sim 0.2-0.7$ assuming the empirical relation of $R_{\rm NSC}=8.3~{\rm pc}(M_{\rm SMBH}/3.1\times 10^8~\Msun)^{0.154}$ \citep{Scott13,Kormendy2013,Georgiev16}, 
and last
the inflow rate of an AGN disk in units of the Eddington rate varies within ${\dot M}_{\rm inflow}/{\dot M}_{\rm Edd}\sim0.004-1.2$, 
where ${\dot M}_{\rm Edd}\equiv L_{\rm Edd}/c^2 \eta_{\rm rad}$ with radiative efficiency of $\eta_{\rm rad}=0.1$, and 
$L_{\rm Edd}$ is the Eddington luminosity of the SMBH.

We find that the optical luminosity of the flare is lower than the kinetic power of the jet ($f_{\rm op/kin}<1$) for all the pairs (Table~\ref{table:properties_model}). 
Considering the bolometric correction of the breakout emission in the optical band ($\sim 10$, Paper~I), our scenario demands $f_{\rm op/kin}\lesssim 0.1$. For the pairs that do not satisfy this condition, we need to use $f_{\rm jet/BHL} > 10$. Note that a higher value ($f_{\rm jet/BHL}\lesssim 100$) is possible by considering the enhancement of the accretion rate due to shocks caused by recoil kicks at merger (Appendix.~B of Paper~I).

In the right panel of Fig.~\ref{fig:sed}, 
the SEDs for thermal emission from the breakout of the jet head for Pairs~1--12 are displayed. 
As they are bright at $\sim 10^{17}$--$10^{19}~{\rm Hz}$ and the duration is long ($\gtrsim 10^6~{\rm s}$), 
they can be observed by X-ray telescopes, such as 
the {\it Swift} X-ray telescope (XRT, \citealt{Burrows2005}), 
Chandra, XMM Newton \citep{Jansen2001}, 
and the Nuclear Spectroscopic Telescope Array (NuSTAR, \citealt{Harrison2013}). 
Note that the shock cooling emission associated with the breakout emission 
in all the pairs 
is so dim ($\ll 10^{40}~{\rm erg/s}$) that it is buried within the AGN variability.

The derived values of $R_{\rm BH}/R_{\rm NSC}\sim 0.2-0.7\sim 1-8$~pc are 
much larger than in the shock-cooling case.  
The different locations are driven by the phases of the emission. 
In our model, the duration of flares corresponds to the timescale at which the diffusion timescale is equal to the dynamical timescale (e.g.~Eqs.~ \ref{eq:rho_diff} and \ref{eq:t_bo} for the shock cooling and breakout emission cases, respectively). 
This timescale is inversely proportional to the density of the ejecta or of the AGN disk, as well as to the square of the velocity of the ejecta or of the shocks in the shock cooling and breakout emission scenarios, respectively. 
As a result, 
since the velocities are typically comparable (Tables~\ref{table:properties_cooling} and \ref{table:properties_model}), 
the density of the AGN disk in the breakout emission scenario
must be similar to
the density of the ejecta in the shock cooling emission scenario, in order to reproduce the duration of the flares. Additionally, 
in the shock cooling emission scenario the density of the ejecta is much lower than the local AGN density due to adiabatic expansion (Eq.~\ref{eq:rho_zero}). 
Thus, the AGN density is typically much lower in the breakout emission scenario, 
and therefore the distance from the SMBH needs to be larger. 
These larger distance are 
consistent with the scenario that BHs in nuclear star clusters are captured by AGN disks and merge with one other. Here, the migration timescale of objects in AGN disks around massive SMBHs as inferred for the flaring AGNs (Table~\ref{table:properties_events}) is so long that migration before BH mergers is less expected.

To check whether the values derived for the aspect ratio ($h_{\rm AGN}$) are plausible, 
we calculate the aspect ratio of the AGN disk ($h_{\rm AGN,TQM}$) adopting the model in \citet{Thompson05}, using $R_{\rm BH}$, $M_{\rm SMBH}$, ${\dot M}_{\rm inflow}$, and $m$. 
We find that the values of $h_{\rm AGN,TQM}$ are roughly comparable to $h_{\rm AGN}$ within a factor of $\sim 2$ (9th column of Table~\ref{table:properties_model}). 
Such moderate differences could arise due to the variation of $f_{\rm corr}$ reflecting the variation of the inclination of jets. 
In Table~\ref{table:properties_model}, we list $f_{\rm corr}$ at which $h_{\rm AGN,TQM}=h_{\rm AGN}$. 
This shows that $h_{\rm AGN,TQM}=h_{\rm AGN}$ is satisfied by reasonable values for $f_{\rm corr}$ around $\sim 3$--$5$.

By comparing $L_{\rm AGN}/L_{\rm Edd}$ in Table~\ref{table:properties_events} and ${\dot M}_{\rm inflow}/{\dot M}_{\rm Edd}$ in Table~\ref{table:properties_model}, 
we can determine the value of $m$ at which 
$L_{\rm AGN}/L_{\rm Edd}= {\dot M}_{\rm inflow}/{\dot M}_{\rm Edd}$ as ${\dot M}_{\rm inflow}\propto m$. 
For the pairs~11 and 12, $m\geq 3$ is required to satisfy $L_{\rm AGN}/L_{\rm Edd}\leq {\dot M}_{\rm inflow}/{\dot M}_{\rm Edd}$. 
In the other pairs (1--10), $m$ can be as low as $\geq 0.003$--$0.5$ to satisfy this condition. 
We presume that $m\lesssim 1$ is acceptable \citep[e.g.][]{Hopkins2010,Begelman2023,Begelman2023b}, although there are no studies constraining the possible ranges of $m$ as far as we know. 
Note that the inflow rate 
at pc scales 
can be much higher than the accretion rate onto the central SMBH \citep[e.g.][]{Izumi2023}, since a large fraction of gas is possibly consumed by star formation and outflows, 
as predicted in \citet{Blandford1999}, \citet{Thompson05}, and \citet{Collin2008}. 
Then ${\dot M}_{\rm inflow}/{\dot M}_{\rm Edd}$ is allowed to be much larger than $L_{\rm AGN}/L_{\rm Edd}$, and the required value of $m$ is enhanced. 
Hence, 
pairs~11 and 12 are not well explained by  breakout emission  due to the required high values of $m$. 
On the other hand, it is not clear whether AGN disks are usually in steady state, which might complicate the comparison between the inflow rate and the AGN luminosity.

\section{Tests of the Model}

\label{sec:discussion}

In the following we discuss how our model can be tested by examining the distribution expected for the observable properties. 
From \S~\ref{sec:results_cooling} and ~\ref{sec:results_breakout} we see that 
values for each model parameter are distributed in a narrow range (Tables~\ref{table:properties_cooling} and \ref{table:properties_model}). 
We discuss  
whether this is because flares originating from merging BHs tend to have well-defined characteristic properties, or because of observational selection effects in the way the search was conducted by \citet{Graham2023}. 
Flares were searched with specific ranges of parameters, and in particular with the rise time within 5--100$~{\rm day}$, the decay time within 10--200$~{\rm day}$, and the delay time from the GW detection in the interval 0--200$~{\rm day}$. 

In order for electromagnetic flares to be found in association with BH mergers, BHs 
need to typically merge in bright AGNs. 
This is because most BH mergers reported by LIGO/Virgo/KAGRA are found to merge at luminosity distances of several Gpc \citep{LIGO20_O3_Catalog}. 
At such large distances, AGNs are easily missed unless they are bright. 
Indeed, the host 
SMBH masses for the flares reported by \citet{Graham2023} are $\gtrsim 10^{8}~\Msun$, 
hence so massive that AGNs are rarely missed in AGN searches at the distances of the GW events. 
Also, assuming a luminosity distance of $d_{\rm L}\sim 3~{\rm Gpc}$, SMBH mass of $10^8~\Msun$, Eddington ratio of $\sim 1$, bolometric correction of 5, and the fraction of the variable luminosity compared to the average luminosity $\sim 0.1$, the flux of variable flares in the AGN is $\sim 2\times 10^{-13}~{\rm erg/s/cm^2}$, which falls just above the sensitivity of ZTF $\sim 10^{-13}~{\rm erg/s/cm^2}$. 
Conversely, it is difficult to find flares associated with BH mergers occurring in AGNs around less massive SMBHs through surveys for AGN variability.

We next estimate the detection rate of electromagnetic flares associated with BH mergers based on our scenario. 
If BH mergers actually produce electromagnetic flares as found in \citet{Graham2023}, 
the rate of such flares is comparable to or less than the rate of BH mergers ($\sim O(10)~{\rm Gpc^{-3} yr^{-1}}$). If flares within $\lesssim 3~{\rm Gpc}$ are observable by the ZTF and all merging BHs produce electromagnetic flares, up to $N_{\rm BBH,3Gpc}\sim 300$ flares associated with BH mergers can be detected per year. 
On the other hand, we have estimated that a small fraction of BH mergers ($f_{\rm EM/BBH}\lesssim 0.02$) could accompany detectable electromagnetic flares in our model (see \S~4.1 in Paper~I for discussions). 
Then, the seven flares discovered by \citep{Graham2023} is roughly consistent with the number of flares expected during LIGO/Virgo/KAGRA~O3, which is estimated as 
$N_{\rm EM/GW,O3}\sim N_{\rm BBH,3Gpc}f_{\rm EM/BBH}t_{\rm O3}\lesssim 10$, where $t_{\rm O3}\sim 1.1$ is the duration of the O3 operation in the unit of year. 
Constraints on the frequency of flares, whose properties can be explained by emission from BHs, are hence useful to test 
our model. 
Here, note that dimmer more frequent flares would be contributed by solitary BHs \citep{Tagawa2023_solitary}.

\subsection{Shock cooling emission}

\label{sec:dis_cooling}

From Table~\ref{table:properties_cooling}, we note that
there are moderate variations in $R_{\rm BH}$ and 
$T_{\rm BB}$ in the shock cooling emission scenario. 
To consider a possible test of this model, we discuss its dependence on the observed properties, and the range of these properties for which the shock cooling emission scenario 
is inconsistent. 
If there are events with high or low $T_{\rm BB}$, 
extremely large or small $R_{\rm BH}$, 
or $R_{\rm BO}/R_{\rm diff}\geq 1$, 
such flares are inconsistent with the shock cooling emission scenario. 
However, the dependence of the variables $T_{\rm BB}$, $R_{\rm BH}$, $R_{\rm BO}/R_{\rm diff}$ on the ratio of the observed parameters ($L_{\rm AGN}$, $L_{\rm obs}$, $m_{\rm BH}$, $M_{\rm SMBH}$, and $t_{\rm diff}$) over the adjustment parameters ($f_{\rm corr}$ and $f_{\rm jet/BHL}$) is similar to the dependence of $H_{\rm AGN}/R_{\rm BH}$ on the same ratio. 
For example, for $\beta_{\rm c}/\theta_0<1$, 
$H_{\rm AGN}/R_{\rm BH}$ depends on the observed properties as 
\begin{align}
\label{eq:hr_cooling}
\frac{H_{\rm AGN}}{R_{\rm BH}}=
\frac{9^{4/3} 
L_{\rm AGN}^2 L_{\rm obs}^5 \kappa_{\rm ej}^7 \theta_0^{2/3} f_{\rm corr}^{9}
}
{2^{4/3} 4^4 \pi^{7} f_{\rm jet/BHL}^4
f_{\rm cons}^2 c^{19} 
\alpha^2 \eta_{\rm rad}^2
A^{4/3} G^3}
\nonumber\\
\times \frac{1}{
m_{\rm BH}^{8/3}M_{\rm SMBH}^{1/3}t_{\rm diff}^{16/3} 
}
\nonumber\\
\propto 
L_{\rm obs}^{5}
L_{\rm AGN}^{2}
t_{\rm delay}^{-16/3}
m_{\rm BH}^{-8/3}
f_{\rm corr}^{9}
f_{\rm jet/BHL}^{-4}
,
\end{align}
and $T_{\rm BB}$ depends as 
\begin{align}\label{eq:tbb_cooling}T_{\rm BB}=\frac{9^{3/8} L_{\rm AGN}^{3/8} L_{\rm obs}^{11/8} \kappa_{\rm ej}^{3/2} f_{\rm corr}^{9/4}}{(4\pi)^{1/3} (32\pi^4)^{3/8} f_{\rm jet/BHL}^{9/8}f_{\rm cons}^{3/8} c^{9/2}A^{3/8} G^{3/4}}\nonumber\\\times \frac{1}{ \alpha^{3/8} \eta_{\rm rad}^{3/8}m_{\rm BH}^{3/4}t_{\rm diff}^{5/4}a^{3/8} }\nonumber\\\propto 
L_{\rm obs}^{11/8}L_{\rm AGN}^{3/8}t_{\rm delay}^{-5/4}m_{\rm BH}^{-3/4}
f_{\rm corr}^{9/4}
f_{\rm jet/BHL}^{-9/8}
. 
\end{align}
Then, if we adjust $f_{\rm corr}$ and $f_{\rm jet/BHL}$ so that $H_{\rm AGN}/R_{BH}$ is consistent with a Shakura-Sunyaev disk, 
these parameters ($T_{\rm BB}$, $R_{\rm BH}$, and $R_{\rm BO}/R_{\rm diff}$) also fall in a range of  possible values (similar values to those derived in Table~\ref{table:properties_cooling}), even if events with wide ranges of observed parameters ($L_{\rm AGN}$, $L_{\rm obs}$, $m_{\rm BH}$, $M_{\rm SMBH}$, and $t_{\rm diff}$) are observed. 
Thus, in this model, once $H_{\rm AGN}/R_{\rm BH}$ is adjusted to the value expected in the Shakura-Sunyaev model, $T_{\rm BB}$, $R_{\rm BH}$, and $R_{\rm BO}/R_{\rm diff}$ are then characterized by 
realistic 
values, and the distribution of the observables becomes difficult to use as a further test of the model.

An interesting test of the model is the correlation between the delay time and the duration of the flare. 
This is because the delay time 
is comparable to
the duration of a flare 
as long as $t_{\rm diff}\gg t_{\rm break}$, which is satisfied in pairs 1--12 (Table~\ref{table:properties_cooling}). 
To constrain the correlation coefficient e.g. with uncertainty of $\lesssim 0.3$ by 95 percentile, more than $\gtrsim 50$ events are needed to be observed. 
Hence, more events 
will be very helpful as further diagnostics.
Note that half of the pairs reported by \citet{Graham2023} are unreal as a flare is associated with a few GW events, and hence, we need to derive the correlation excluding the influence from false associations. 
Also, we derived the delay time using {\it digitizer}. 
In addition, the time dependence of the luminosity assumed by \citet{Graham2023} is different from that expected in the shock cooling emission. Thus, both the delay time and the duration suffer from significant uncertainties.  
If these timescales can be well constrained, 
the model can be tested for each event by 
comparing the delay time and the duration.

Another possible test is the detection of the cocoon breakout emission preceding the shock cooling emission. 
In pairs 1--12, the duration and temperature of the cocoon breakout emission range in the intervals $\sim 10$--$10^5~{\rm s}$ and $0.4$--$5~{\rm keV}$; thus
wide-field X-ray surveys, such as Einstein Probe \citep{Yuan2015} and HiZ-GUNDAM \citep{Yonetoku2020c} will be useful to detect the early breakout emission (left panel of Fig.~\ref{fig:sed}). 
Assuming that the luminosity of the breakout emission is similar in magnitude to the jet power, the detectable distance is estimated to be 
\begin{align}
\label{eq:d_sen}
d_{\rm det}=\left(\frac{L_{\rm jet}}{4\pi F_{\rm sen}}\right)^{1/2}\nonumber\\
\sim 1~{\rm Gpc}
\left(\frac{L_{\rm jet}}{10^{46}~{\rm erg/s}}\right)^{1/2}
\left(\frac{F_{\rm sen}}{10^{-11}~{\rm erg/s/cm^2}}\right)^{-1/2},
\end{align}
where $F_{\rm sen}$ is the sensitivity of the facility. 
Here, the closest distance among the flares discovered by \citet{Graham2023} is $\sim {\rm Gpc}$. 
Since the duration of the cocoon breakout emission is $\sim 10$--$10^5~{\rm s}$ in pairs~1--12 and the sensitivity of 
Einstein Probe \citep{Yuan2015} and HiZ-GUNDAM is 
$F_{\rm sen}\sim 10^{-11}~{\rm erg/s/cm^2}$
for $t_{\rm int}\sim 10^4~{\rm s}$, 
events at the luminosity distance of $\sim {\rm Gpc}$ can be detected 
if $L_{\rm jet}\gtrsim 10^{46}~{\rm erg/s}$. 
If both the breakout emission and the shock cooling emission are detected from the same AGN, it can be a robust test of this model.

It is notable that in the shock cooling model, the color keeps evolving to 
redder 
in all pairs (Fig.~\ref{fig:lc}~d). 
The 
$m_{g} - m_{r}$ 
evolves by $\sim 0.4$ in $50~{\rm days}$. 
Such mild evolution of the color can be a strong test of this model. 
On the other hand, for non-thermal emission from the breakout of the jet head, the color is expected to be unchanged. 
Note that the colors are presented for pair~3 in Fig.~2 of \citet{Graham20}, revealing almost no evolution. However, due to the contamination from the host AGN emission, any color evolution is likely difficult to constrain. 
To derive the color evolution more precisely, observations at additional frequencies, e.g. by ULTRASAT (black lines of Fig.~\ref{fig:lc}~c), would also be useful. 
Hence, when available, the evolution of the color will be an additional diagnostic in future observations.

\subsection{Breakout emission}

\label{sec:dis_breakout}

Next we discuss ways to test the breakout emission scenario. 
To do this, we first present the dependence of physical quantities on observables as 
\begin{align}
\beta_{\rm h} \propto \theta_0^{-12/11} f_{\rm jet/BHL}^{3/11}f_{\rm corr}^{-3/11} t_{\rm delay}^{-3/11} t_{\rm duration}^{1/22}
, \\
\hspace{0.1\baselineskip}\nonumber\\
H_{\rm AGN}\propto \theta_0^{-12/11} f_{\rm jet/BHL}^{3/11}f_{\rm corr}^{-14/11} t_{\rm delay}^{8/11} t_{\rm duration}^{1/22}
, \\
\hspace{0.1\baselineskip}\nonumber\\
\rho_{\rm AGN} \propto \theta_0^{24/11} f_{\rm jet/BHL}^{-6/11}f_{\rm corr}^{6/11} t_{\rm delay}^{6/11} t_{\rm duration}^{-12/22}
,\\
\hspace{0.1\baselineskip}\nonumber\\
R_{\rm BH} \propto \theta_0^{-8/11} f_{\rm jet/BHL}^{2/11}M_{\rm SMBH}^{1/3}f_{\rm corr}^{-2/11} t_{\rm delay}^{-2/11} t_{\rm duration}^{4/22}
,\\
\hspace{0.1\baselineskip}\nonumber\\
L_{\rm j} \propto \theta_0^{8/11} f_{\rm jet/BHL}^{9/11}f_{\rm corr}^{-9/11} t_{\rm delay}^{13/11} t_{\rm duration}^{-19/22}
,\\ 
\hspace{0.1\baselineskip}\nonumber\\
{\dot M}_{\rm inflow} \propto \theta_0^{4/11} f_{\rm jet/BHL}^{-1/11}m M_{\rm SMBH}^{1/2}f_{\rm corr}^{-17/11} t_{\rm delay}^{23/11} t_{\rm duration}^{-13/22}.  
\end{align}

For the breakout emission from the jet head, 
if the shocked gas becomes relativistic ($\beta_{\rm h}\gtrsim 0.8$), 
the probability of observing non-thermal emission is significantly reduced due to 
the relativistic beaming effects, 
the shift of the minimum energy, and 
the time dilution (as shown in Fig.~3~b of Paper~I). 
Thus, if the shock is relativistic, 
the breakout emission is presumably not observed by current facilities, such as ZTF. 
Since $\beta_{\rm h}$ is estimated to be $\sim 0.2$--$0.4$ for the observed flares, if $\beta_{\rm h}$ is higher by a factor of $\sim 3$ compared to our estimates, 
flares cannot be explained by breakout emission. 
$\beta_{\rm h}\gtrsim 0.8$ is satisfied in all the pairs 
if events with delay time of $t_{\rm delay}\lesssim 2~{\rm day}$ are found. 
Note that this is not compensated by $f_{\rm corr}$ and $f_{\rm jet/BHL}$ (as in the shock cooling emission scenario). 
This is because $\beta_{\rm h}$ is reduced only by a factor of $\sim 1.1$ by enhancing $f_{\rm corr}$ to the maximum value ($\sim 1/\theta_0$). Also, if $f_{\rm jet/BHL}$ is reduced to lower $\beta_{\rm h}$, $f_{\rm op/kin}$ becomes larger than 1, which violates another requirement for the model. 
If associations between the flares and GWs are due to random coincidence, the delay time is expected to be distributed uniformly in the range of $0$--$200~{\rm day}$. 
Assuming a uniform distribution, we can test the breakout emission scenario 
at the $\sim 1~\sigma$ and $\sim 2~\sigma$ levels 
after discovering $\gtrsim 100$ and $\gtrsim 300$ events, respectively, 
by checking if there are events with $t_{\rm delay}\lesssim 2~{\rm day}$. 
If we find optical flares with $t_{\rm delay}\lesssim 2~{\rm day}$, the breakout emission scenario 
is disfavored.

If $t_{\rm duration}$ is enhanced by one order of magnitude, 
the jet power is reduced by a similar factor, and the fraction of the optical luminosity to the jet power exceeds one for Pairs 1--3, 6, 11, and 12, and then, the breakout emission scenario 
becomes inconsistent for these pairs. 
However, the enhancement of $L_j$ due to long $t_{\rm duration}$ can be compensated by $f_{\rm jet}$ and $f_{\rm corr}$ up to about a factor of $\sim 10$, which is limited by the requirement for $\beta_{\rm h}$ $(\lesssim 0.8)$ as discussed above. 
Hence, the model is not well tested by $t_{\rm duration}$.

Currently, the properties of the observed flares 
satisfy the conditions required for the breakout emission scenario. 
To be consistent with this scenario, 
the delay time should not be shorter than 
$\sim 2~{\rm day}$, 
and the color of flares should be distributed in a narrow range as discussed in \S~\ref{sec:results_cooling}. 
Thus, to test whether there are flares with properties being inconsistent with 
the breakout emission scenario, 
more events will need to be observed.

\section{Conclusions}

\label{sec:conclusions}

In this paper 
we have presented 
the properties of emission from shocks emerging from collisions between AGN gas and a jet launched from a merger remnant BH in an AGN disk. Our model 
includes the evolution throughout the shock breakout and subsequent cooling emission phases. 
We then applied this model to 
the candidate flares reported in \citet{Graham2023}. 
Our results are summarized as follows.

\begin{enumerate}

\item 
We fit the characteristic features of all of the events with both emission processes. Both processes could fit each observation, suggesting that such fits may themselves be insufficient to rule out the AGN origin for the flares. 
The reconstructed parameters might then be indicative of the selection algorithm determining the false alarm rate of associations. 

\item 
While both processes could be made consistent with the observed events with appropriate parameter selection, we found that the implied merger distance from the central SMBH is markedly different for the two processes. Specifically, shock cooling emission can explain the observed properties if the mergers happen $R_{\rm BH}\sim 0.06$--$0.3~{\rm pc}$ from the SMBH, which is consistent with the locations of AGN-assisted merger models (e.g. \citealt{Tagawa19}). 
On the other hand, breakout emission would require a much larger distance of $R_{\rm BH}\sim 1$--$8~{\rm pc}$ to explain the observed flare duration. This may be possible for mergers in AGNs with high SMBH masses, in which migration is inefficient (e.g. \citealt{Perna2021_AICs}). 

\item 
Follow-up observations could help further constrain the reconstructed parameters of the events. X-ray observations would have to be made prior to the optical detection, which 
likely requires future wide-field surveys. 
In addition, follow-up observations determining the spectral evolution of the electromagnetic flares would be important.

\end{enumerate}

\acknowledgments

H.T. was supported by the National Key R\&D Program of China (Grant No. 2021YFC2203002) and the National Natural Science Foundation of China (Grant No. 12173071). 
S.S.K. was supported by 
Japan Society for the Promotion of Science (JSPS) KAKENHI 
grant Number 22K14028, 21H04487, and 23H04899, and 
the Tohoku Initiative for Fostering Global Researchers for Interdisciplinary Sciences (TI-FRIS) of MEXT's Strategic Professional Development Program for Young Researchers. 
Z.H. was supported by NASA grant 80NSSC22K0822 and NSF grant AST-2006176. 
R.P. acknowledges support by NSF award AST-2006839. 
I.B. acknowledges the support of the Alfred P. Sloan Foundation and NSF grants PHY-1911796 and PHY-2110060.

\iffalse ... 
\fi

\bibliographystyle{aasjournal}
\bibliography{agn_bhm}

\begin{thebibliography}{}
\expandafter\ifx\csname natexlab\endcsname\relax\def\natexlab#1{#1}\fi
\providecommand{\url}[1]{\href{#1}{#1}}
\providecommand{\dodoi}[1]{doi:~\href{http://doi.org/#1}{\nolinkurl{#1}}}
\providecommand{\doeprint}[1]{\href{http://ascl.net/#1}{\nolinkurl{http://ascl.net/#1}}}
\providecommand{\doarXiv}[1]{\href{https://arxiv.org/abs/#1}{\nolinkurl{https://arxiv.org/abs/#1}}}

\bibitem[{{Aasi} {et~al.}(2015)}]{2015CQGra..32g4001L}
{Aasi}, J., {et~al.} 2015, \cqg, 32, 074001,
  \dodoi{10.1088/0264-9381/32/7/074001}

\bibitem[{{Abbott} {et~al.}(2020){Abbott}, {Abbott}, {Abraham}, {Acernese},
  {Ackley}, {Adams}, {Adams}, {Adhikari}, {Adya}, {Affeldt}, {Agathos},
  {Agatsuma}, {Aggarwal}, {Aguiar}, {Aiello}, {Ain}, \&
  {Ajith}}]{LIGO20_O3_Catalog}
{Abbott}, R., {Abbott}, T.~D., {Abraham}, S., {et~al.} 2020, arXiv e-prints,
  arXiv:2010.14527.
\newblock \doarXiv{2010.14527}

\bibitem[{{Acernese} {et~al.}(2015){Acernese}, {Agathos}, {Agatsuma}, {Aisa},
  {Allemandou}, {Allocca}, {Amarni}, {Astone}, {Balestri}, \&
  {Ballardin}}]{2015CQGra..32b4001A}
{Acernese}, F., {Agathos}, M., {Agatsuma}, K., {et~al.} 2015, Classical and
  Quantum Gravity, 32, 024001, \dodoi{10.1088/0264-9381/32/2/024001}

\bibitem[{{Akutsu} {et~al.}(2021){Akutsu}, {Ando}, {Arai}, {Arai}, {Araki},
  {Araya}, {Aritomi}, {Aso}, {Bae}, {Bae}, {et~al.}}]{2021PTEP.2021eA101A}
{Akutsu}, T., {Ando}, M., {Arai}, K., {et~al.} 2021, Progress of Theoretical
  and Experimental Physics, 2021, 05A101, \dodoi{10.1093/ptep/ptaa125}

\bibitem[{{Antonini} {et~al.}(2022){Antonini}, {Gieles}, {Dosopoulou}, \&
  {Chattopadhyay}}]{Antonini2022}
{Antonini}, F., {Gieles}, M., {Dosopoulou}, F., \& {Chattopadhyay}, D. 2022,
  arXiv e-prints, arXiv:2208.01081, \dodoi{10.48550/arXiv.2208.01081}

\bibitem[{{Antonini} {et~al.}(2017){Antonini}, {Toonen}, \&
  {Hamers}}]{Antonini17}
{Antonini}, F., {Toonen}, S., \& {Hamers}, A.~S. 2017, \apj, 841, 77,
  \dodoi{10.3847/1538-4357/aa6f5e}

\bibitem[{{Arca Sedda} {et~al.}(2023){Arca Sedda}, {Naoz}, \&
  {Kocsis}}]{ArcaSedda2023}
{Arca Sedda}, M., {Naoz}, S., \& {Kocsis}, B. 2023, arXiv e-prints,
  arXiv:2302.14071, \dodoi{10.48550/arXiv.2302.14071}

\bibitem[{{Arnett}(1980)}]{Arnett1980}
{Arnett}, W.~D. 1980, \apj, 237, 541, \dodoi{10.1086/157898}

\bibitem[{{Ashton} {et~al.}(2020){Ashton}, {Ackley}, {Maga{\~n}a Hernandez}, \&
  {Piotrzkowski}}]{Ashton2020}
{Ashton}, G., {Ackley}, K., {Maga{\~n}a Hernandez}, I., \& {Piotrzkowski}, B.
  2020, arXiv e-prints, arXiv:2009.12346.
\newblock \doarXiv{2009.12346}

\bibitem[{{Bagoly} {et~al.}(2016){Bagoly}, {Sz{\'e}csi}, {Bal{\'a}zs},
  {Csabai}, {Horv{\'a}th}, {Dobos}, {Lichtenberger}, \&
  {T{\'o}th}}]{Bagoly2016}
{Bagoly}, Z., {Sz{\'e}csi}, D., {Bal{\'a}zs}, L.~G., {et~al.} 2016, \aap, 593,
  L10, \dodoi{10.1051/0004-6361/201628569}

\bibitem[{{Banerjee}(2017)}]{Banerjee17}
{Banerjee}, S. 2017, \mnras, 467, 524.
\newblock \url{http://adsabs.harvard.edu/abs/2017MNRAS.467..524B}

\bibitem[{{Barack} {et~al.}(2019){Barack}, {Cardoso}, {Nissanke}, {Sotiriou},
  {Askar}, {Belczynski}, {Bertone}, {Bon}, {Blas}, {Brito}, {Bulik}, {Burrage},
  {Byrnes}, {Caprini}, {Chernyakova}, {Chru{\'s}ciel}, {Colpi}, {Ferrari},
  {Gaggero}, {Gair}, {Garc{\'\i}a-Bellido}, {Hassan}, {Heisenberg}, {Hendry},
  {Heng}, {Herdeiro}, {Hinderer}, {Horesh}, {Kavanagh}, {Kocsis}, {Kramer}, {Le
  Tiec}, {Mingarelli}, {Nardini}, {Nelemans}, {Palenzuela}, {Pani}, {Perego},
  {Porter}, {Rossi}, {Schmidt}, {Sesana}, {Sperhake}, {Stamerra}, {Stein},
  {Tamanini}, {Tauris}, {Urena-L{\'o}pez}, {Vincent}, {Volonteri}, {Wardell},
  {Wex}, {Yagi}, {Abdelsalhin}, {Aloy}, {Amaro-Seoane}, {Annulli},
  {Arca-Sedda}, {Bah}, {Barausse}, {Barakovic}, {Benkel}, {Bennett}, {Bernard},
  {Bernuzzi}, {Berry}, {Berti}, {Bezares}, {Juan Blanco-Pillado},
  {Bl{\'a}zquez-Salcedo}, {Bonetti}, {Bo{\v{s}}kovi{\'c}}, {Bosnjak},
  {Bricman}, {Br{\"u}gmann}, {Capelo}, {Carloni}, {Cerd{\'a}-Dur{\'a}n},
  {Charmousis}, {Chaty}, {Clerici}, {Coates}, {Colleoni}, {Collodel},
  {Comp{\`e}re}, {Cook}, {Cordero-Carri{\'o}n}, {Correia}, {de la
  Cruz-Dombriz}, {Czinner}, {Destounis}, {Dialektopoulos}, {Doneva}, {Dotti},
  {Drew}, {Eckner}, {Edholm}, {Emparan}, {Erdem}, {Ferreira}, {Ferreira},
  {Finch}, {Font}, {Franchini}, {Fransen}, {Gal'tsov}, {Ganguly}, {Gerosa},
  {Glampedakis}, {Gomboc}, {Goobar}, {Gualtieri}, {Guendelman}, {Haardt},
  {Harmark}, {Hejda}, {Hertog}, {Hopper}, {Husa}, {Ihanec}, {Ikeda}, {Jaodand},
  {Jetzer}, {Jimenez-Forteza}, {Kamionkowski}, {Kaplan}, {Kazantzidis},
  {Kimura}, {Kobayashi}, {Kokkotas}, {Krolik}, {Kunz}, {L{\"a}mmerzahl},
  {Lasky}, {Lemos}, {Levi Said}, {Liberati}, {Lopes}, {Luna}, {Ma}, {Maggio},
  {Mangiagli}, {Martinez Montero}, {Maselli}, {Mayer}, {Mazumdar}, {Messenger},
  {M{\'e}nard}, {Minamitsuji}, {Moore}, {Mota}, {Nampalliwar}, {Nerozzi},
  {Nichols}, {Nissimov}, {Obergaulinger}, {Obers}, {Oliveri}, {Pappas},
  {Pasic}, {Peiris}, {Petrushevska}, {Pollney}, {Pratten}, {Rakic}, {Racz},
  {Radia}, {Ramazano{\u{g}}lu}, {Ramos-Buades}, {Raposo}, {Rogatko},
  {Rosca-Mead}, {Rosinska}, {Rosswog}, {Ruiz-Morales}, {Sakellariadou},
  {Sanchis-Gual}, {Sharan Salafia}, {Samajdar}, {Sintes}, {Smole}, {Sopuerta},
  {Souza-Lima}, {Stalevski}, {Stergioulas}, {Stevens}, {Tamfal},
  {Torres-Forn{\'e}}, {Tsygankov}, {{\"U}nl{\"u}t{\"u}rk}, {Valiante}, {van de
  Meent}, {Velhinho}, {Verbin}, {Vercnocke}, {Vernieri}, {Vicente},
  {Vitagliano}, {Weltman}, {Whiting}, {Williamson}, {Witek}, {Wojnar}, {Yakut},
  {Yan}, {Yazadjiev}, {Zaharijas}, \& {Zilh{\~a}o}}]{Barack2019}
{Barack}, L., {Cardoso}, V., {Nissanke}, S., {et~al.} 2019, Classical and
  Quantum Gravity, 36, 143001, \dodoi{10.1088/1361-6382/ab0587}

\bibitem[{{Bartos} {et~al.}(2017){Bartos}, {Kocsis}, {Haiman}, \&
  {M{\'a}rka}}]{Bartos17}
{Bartos}, I., {Kocsis}, B., {Haiman}, Z., \& {M{\'a}rka}, S. 2017, \apj, 835,
  165.
\newblock \url{http://adsabs.harvard.edu/abs/2017ApJ...835..165B}

\bibitem[{{Bartos} {et~al.}(2023){Bartos}, {Rosswog}, {Gayathri}, {Miller},
  {Veske}, \& {Marka}}]{2023arXiv230210350B}
{Bartos}, I., {Rosswog}, S., {Gayathri}, V., {et~al.} 2023, arXiv e-prints,
  arXiv:2302.10350, \dodoi{10.48550/arXiv.2302.10350}

\bibitem[{{Begelman} \& {Armitage}(2023)}]{Begelman2023}
{Begelman}, M.~C., \& {Armitage}, P.~J. 2023, \mnras, 521, 5952,
  \dodoi{10.1093/mnras/stad914}

\bibitem[{{Begelman} \& {Silk}(2023)}]{Begelman2023b}
{Begelman}, M.~C., \& {Silk}, J. 2023, arXiv e-prints, arXiv:2305.19081,
  \dodoi{10.48550/arXiv.2305.19081}

\bibitem[{{Belczynski} {et~al.}(2016){Belczynski}, {Daniel}, {Bulik}, \&
  {O'Shaughnessy}}]{Belczynski16}
{Belczynski}, K., {Daniel}, E.~H., {Bulik}, T., \& {O'Shaughnessy}, R. 2016,
  \nat, 534, 512.
\newblock \url{http://adsabs.harvard.edu/abs/2016Natur.534..512B}

\bibitem[{{Bellovary} {et~al.}(2016){Bellovary}, {Mac Low}, {McKernan}, \&
  {Ford}}]{Bellovary16}
{Bellovary}, J.~M., {Mac Low}, M.~M., {McKernan}, B., \& {Ford}, K.~E.~S. 2016,
  \apj, 819, L17.
\newblock \url{http://adsabs.harvard.edu/abs/2016ApJ...819L..17B}

\bibitem[{{Berger}(2014)}]{Berger2014}
{Berger}, E. 2014, \araa, 52, 43, \dodoi{10.1146/annurev-astro-081913-035926}

\bibitem[{{Blandford} \& {Begelman}(1999)}]{Blandford1999}
{Blandford}, R.~D., \& {Begelman}, M.~C. 1999, \mnras, 303, L1,
  \dodoi{10.1046/j.1365-8711.1999.02358.x}

\bibitem[{{Blandford} \& {Znajek}(1977)}]{Blandford1977}
{Blandford}, R.~D., \& {Znajek}, R.~L. 1977, \mnras, 179, 433,
  \dodoi{10.1093/mnras/179.3.433}

\bibitem[{{Boekholt} {et~al.}(2023){Boekholt}, {Rowan}, \&
  {Kocsis}}]{Boekholt2023}
{Boekholt}, T. C.~N., {Rowan}, C., \& {Kocsis}, B. 2023, \mnras, 518, 5653,
  \dodoi{10.1093/mnras/stac3495}

\bibitem[{{Bromberg} {et~al.}(2011){Bromberg}, {Nakar}, {Piran}, \&
  {Sari}}]{Bromberg2011}
{Bromberg}, O., {Nakar}, E., {Piran}, T., \& {Sari}, R. 2011, \apj, 740, 100,
  \dodoi{10.1088/0004-637X/740/2/100}

\bibitem[{{Burrows} {et~al.}(2005){Burrows}, {Hill}, {Nousek}, {Kennea},
  {Wells}, {Osborne}, {Abbey}, {Beardmore}, {Mukerjee}, {Short}, {Chincarini},
  {Campana}, {Citterio}, {Moretti}, {Pagani}, {Tagliaferri}, {Giommi},
  {Capalbi}, {Tamburelli}, {Angelini}, {Cusumano}, {Br{\"a}uninger}, {Burkert},
  \& {Hartner}}]{Burrows2005}
{Burrows}, D.~N., {Hill}, J.~E., {Nousek}, J.~A., {et~al.} 2005, \ssr, 120,
  165, \dodoi{10.1007/s11214-005-5097-2}

\bibitem[{{Calder{\'o}n Bustillo} {et~al.}(2021){Calder{\'o}n Bustillo},
  {Leong}, {Chandra}, {McKernan}, \& {Ford}}]{CalderonBustillo21}
{Calder{\'o}n Bustillo}, J., {Leong}, S. H.~W., {Chandra}, K., {McKernan}, B.,
  \& {Ford}, K.~E.~S. 2021, arXiv e-prints, arXiv:2112.12481.
\newblock \doarXiv{2112.12481}

\bibitem[{{Chen} {et~al.}(2023{\natexlab{a}}){Chen}, {Ren}, \&
  {Dai}}]{Chen2023}
{Chen}, K., {Ren}, J., \& {Dai}, Z.-G. 2023{\natexlab{a}}, \apj, 948, 136,
  \dodoi{10.3847/1538-4357/acc45f}

\bibitem[{{Chen} {et~al.}(2023{\natexlab{b}}){Chen}, {Jiang}, {Goodman}, \&
  {Ostriker}}]{ChenY2023}
{Chen}, Y.-X., {Jiang}, Y.-F., {Goodman}, J., \& {Ostriker}, E.~C.
  2023{\natexlab{b}}, \apj, 948, 120, \dodoi{10.3847/1538-4357/acc023}

\bibitem[{{Choksi} {et~al.}(2023){Choksi}, {Chiang}, {Fung}, \&
  {Zhu}}]{Choksi2023}
{Choksi}, N., {Chiang}, E., {Fung}, J., \& {Zhu}, Z. 2023, \mnras,
  \dodoi{10.1093/mnras/stad2269}

\bibitem[{{Collin} \& {Zahn}(2008)}]{Collin2008}
{Collin}, S., \& {Zahn}, J.~P. 2008, \aap, 477, 419,
  \dodoi{10.1051/0004-6361:20078191}

\bibitem[{{Connaughton} {et~al.}(2016){Connaughton}, {Burns}, {Goldstein},
  {Blackburn}, {Briggs}, {Zhang}, {Camp}, {Christensen}, {Hui}, {Jenke},
  {Littenberg}, {McEnery}, {Racusin}, {Shawhan}, {Singer}, {Veitch},
  {Wilson-Hodge}, {Bhat}, {Bissaldi}, {Cleveland}, {Fitzpatrick}, {Giles},
  {Gibby}, {von Kienlin}, {Kippen}, {McBreen}, {Mailyan}, {Meegan}, {Paciesas},
  {Preece}, {Roberts}, {Sparke}, {Stanbro}, {Toelge}, \&
  {Veres}}]{Connaughton2016}
{Connaughton}, V., {Burns}, E., {Goldstein}, A., {et~al.} 2016, \apjl, 826, L6,
  \dodoi{10.3847/2041-8205/826/1/L6}

\bibitem[{{de Mink} \& {King}(2017)}]{deMink2017}
{de Mink}, S.~E., \& {King}, A. 2017, \apjl, 839, L7,
  \dodoi{10.3847/2041-8213/aa67f3}

\bibitem[{{DeLaurentiis} {et~al.}(2023){DeLaurentiis}, {Haiman}, \&
  {Epstein-Martin}}]{DeLaurentiis2023}
{DeLaurentiis}, S., {Haiman}, Z., \& {Epstein-Martin}, M. 2023, in American
  Astronomical Society Meeting Abstracts, Vol.~55, American Astronomical
  Society Meeting Abstracts, 351.02

\bibitem[{{Derdzinski} {et~al.}(2021){Derdzinski}, {D'Orazio}, {Duffell},
  {Haiman}, \& {MacFadyen}}]{Derdzinski21}
{Derdzinski}, A., {D'Orazio}, D., {Duffell}, P., {Haiman}, Z., \& {MacFadyen},
  A. 2021, \mnras, 501, 3540, \dodoi{10.1093/mnras/staa3976}

\bibitem[{{Derdzinski} \& {Mayer}(2023)}]{Derdzinski23}
{Derdzinski}, A., \& {Mayer}, L. 2023, \mnras, 521, 4522,
  \dodoi{10.1093/mnras/stad749}

\bibitem[{{Derdzinski} {et~al.}(2019){Derdzinski}, {D'Orazio}, {Duffell},
  {Haiman}, \& {MacFadyen}}]{Derdzinski19}
{Derdzinski}, A.~M., {D'Orazio}, D., {Duffell}, P., {Haiman}, Z., \&
  {MacFadyen}, A. 2019, \mnras, 486, 2754, \dodoi{10.1093/mnras/stz1026}

\bibitem[{{Dominik} {et~al.}(2012){Dominik}, {Belczynski}, {Fryer}, {Holz},
  {Berti}, {Bulik}, {Mandel}, \& {O'Shughnessy}}]{Dominik12}
{Dominik}, M., {Belczynski}, K., {Fryer}, C., {et~al.} 2012, \apj, 759, 52.
\newblock \url{http://adsabs.harvard.edu/abs/2012ApJ...759...52D}

\bibitem[{{Duras} {et~al.}(2020){Duras}, {Bongiorno}, {Ricci}, {Piconcelli},
  {Shankar}, {Lusso}, {Bianchi}, {Fiore}, {Maiolino}, {Marconi}, {Onori},
  {Sani}, {Schneider}, {Vignali}, \& {La Franca}}]{Duras2020}
{Duras}, F., {Bongiorno}, A., {Ricci}, F., {et~al.} 2020, \aap, 636, A73,
  \dodoi{10.1051/0004-6361/201936817}

\bibitem[{{Ford} \& {McKernan}(2022)}]{Ford2022}
{Ford}, K.~E.~S., \& {McKernan}, B. 2022, \mnras, 517, 5827,
  \dodoi{10.1093/mnras/stac2861}

\bibitem[{{Fragione} \& {Kocsis}(2018)}]{Fragione18_EvolvingGC}
{Fragione}, G., \& {Kocsis}, B. 2018, \prl, 121, 161103,
  \dodoi{10.1103/PhysRevLett.121.161103}

\bibitem[{{Fragione} \& {Kocsis}(2019)}]{Fragione19}
---. 2019, \mnras, 486, 4781, \dodoi{10.1093/mnras/stz1175}

\bibitem[{{Gayathri} {et~al.}(2023){Gayathri}, {Wysocki}, {Yang},
  {Shaughnessy}, {Haiman}, {Tagawa}, \& {Bartos}}]{Gayathri2023}
{Gayathri}, V., {Wysocki}, D., {Yang}, Y., {et~al.} 2023, arXiv e-prints,
  arXiv:2301.04187, \dodoi{10.48550/arXiv.2301.04187}

\bibitem[{{Gayathri} {et~al.}(2021){Gayathri}, {Yang}, {Tagawa}, {Haiman}, \&
  {Bartos}}]{2021ApJ...920L..42G}
{Gayathri}, V., {Yang}, Y., {Tagawa}, H., {Haiman}, Z., \& {Bartos}, I. 2021,
  \apjl, 920, L42, \dodoi{10.3847/2041-8213/ac2cc1}

\bibitem[{{Gayathri} {et~al.}(2022){Gayathri}, {Healy}, {Lange}, {O'Brien},
  {Szczepa{\'n}czyk}, {Bartos}, {Campanelli}, {Klimenko}, {Lousto}, \&
  {O'Shaughnessy}}]{Gayathri2022}
{Gayathri}, V., {Healy}, J., {Lange}, J., {et~al.} 2022, Nature Astronomy, 6,
  344, \dodoi{10.1038/s41550-021-01568-w}

\bibitem[{{Generozov} \& {Perets}(2022)}]{Generozov2022}
{Generozov}, A., \& {Perets}, H.~B. 2022, arXiv e-prints, arXiv:2212.11301,
  \dodoi{10.48550/arXiv.2212.11301}

\bibitem[{{Georgiev} {et~al.}(2016){Georgiev}, {Boker}, {Leigh}, {Lutzgendorf},
  \& {Neumayer}}]{Georgiev16}
{Georgiev}, I.~Y., {Boker}, T., {Leigh}, N., {Lutzgendorf}, N., \& {Neumayer},
  N. 2016, \mnras, 457, 2122.
\newblock \url{http://adsabs.harvard.edu/abs/2016MNRAS.457.2122G}

\bibitem[{{Goodman} \& {Tan}(2004)}]{Goodman04}
{Goodman}, J., \& {Tan}, J.~C. 2004, \apj, 608, 108.
\newblock \url{http://adsabs.harvard.edu/abs/2004ApJ...608..108G}

\bibitem[{{Gottlieb} {et~al.}(2018){Gottlieb}, {Nakar}, \&
  {Piran}}]{Gottliev2018}
{Gottlieb}, O., {Nakar}, E., \& {Piran}, T. 2018, \mnras, 473, 576,
  \dodoi{10.1093/mnras/stx2357}

\bibitem[{{Graham} {et~al.}(2020){Graham}, {Ford}, {McKernan}, {Ross}, {Stern},
  {Burdge}, {Coughlin}, {Djorgovski}, {Drake}, {Duev}, {Kasliwal}, {Mahabal},
  {van Velzen}, {Belecki}, {Bellm}, {Burruss}, {Cenko}, {Cunningham}, {Helou},
  {Kulkarni}, {Masci}, {Prince}, {Reiley}, {Rodriguez}, {Rusholme}, {Smith}, \&
  {Soumagnac}}]{Graham20}
{Graham}, M.~J., {Ford}, K.~E.~S., {McKernan}, B., {et~al.} 2020, \prl, 124,
  251102, \dodoi{10.1103/PhysRevLett.124.251102}

\bibitem[{{Graham} {et~al.}(2023){Graham}, {McKernan}, {Ford}, {Stern},
  {Djorgovski}, {Coughlin}, {Burdge}, {Bellm}, {Helou}, {Mahabal}, {Masci},
  {Purdum}, {Rosnet}, \& {Rusholme}}]{Graham2023}
{Graham}, M.~J., {McKernan}, B., {Ford}, K.~E.~S., {et~al.} 2023, \apj, 942,
  99, \dodoi{10.3847/1538-4357/aca480}

\bibitem[{{Grishin} {et~al.}(2021){Grishin}, {Bobrick}, {Hirai}, {Mandel}, \&
  {Perets}}]{Grishin2021}
{Grishin}, E., {Bobrick}, A., {Hirai}, R., {Mandel}, I., \& {Perets}, H.~B.
  2021, \mnras, 507, 156, \dodoi{10.1093/mnras/stab1957}

\bibitem[{{Grishin} {et~al.}(2023){Grishin}, {Gilbaum}, \&
  {Stone}}]{Grishin2023}
{Grishin}, E., {Gilbaum}, S., \& {Stone}, N.~C. 2023, arXiv e-prints,
  arXiv:2307.07546, \dodoi{10.48550/arXiv.2307.07546}

\bibitem[{{Hada} {et~al.}(2013){Hada}, {Kino}, {Doi}, {Nagai}, {Honma},
  {Hagiwara}, {Giroletti}, {Giovannini}, \& {Kawaguchi}}]{Hada2013}
{Hada}, K., {Kino}, M., {Doi}, A., {et~al.} 2013, \apj, 775, 70,
  \dodoi{10.1088/0004-637X/775/1/70}

\bibitem[{{Hada} {et~al.}(2018){Hada}, {Doi}, {Wajima}, {D'Ammando}, {Orienti},
  {Giroletti}, {Giovannini}, {Nakamura}, \& {Asada}}]{Hada2018}
{Hada}, K., {Doi}, A., {Wajima}, K., {et~al.} 2018, \apj, 860, 141,
  \dodoi{10.3847/1538-4357/aac49f}

\bibitem[{{Haiman} {et~al.}(2009){Haiman}, {Kocsis}, \& {Menou}}]{Haiman2009}
{Haiman}, Z., {Kocsis}, B., \& {Menou}, K. 2009, \apj, 700, 1952,
  \dodoi{10.1088/0004-637X/700/2/1952}

\bibitem[{{Hamidani} \& {Ioka}(2023)}]{Humidani2023}
{Hamidani}, H., \& {Ioka}, K. 2023, \mnras, 524, 4841,
  \dodoi{10.1093/mnras/stad1933}

\bibitem[{{Harrison} {et~al.}(2013){Harrison}, {Craig}, {Christensen},
  {Hailey}, {Zhang}, {Boggs}, {Stern}, {Cook}, {Forster}, {Giommi},
  {Grefenstette}, {Kim}, {Kitaguchi}, {Koglin}, {Madsen}, {Mao}, {Miyasaka},
  {Mori}, {Perri}, {Pivovaroff}, {Puccetti}, {Rana}, {Westergaard}, {Willis},
  {Zoglauer}, {An}, {Bachetti}, {Barri{\`e}re}, {Bellm}, {Bhalerao},
  {Brejnholt}, {Fuerst}, {Liebe}, {Markwardt}, {Nynka}, {Vogel}, {Walton},
  {Wik}, {Alexander}, {Cominsky}, {Hornschemeier}, {Hornstrup}, {Kaspi},
  {Madejski}, {Matt}, {Molendi}, {Smith}, {Tomsick}, {Ajello}, {Ballantyne},
  {Balokovi{\'c}}, {Barret}, {Bauer}, {Blandford}, {Brandt}, {Brenneman},
  {Chiang}, {Chakrabarty}, {Chenevez}, {Comastri}, {Dufour}, {Elvis}, {Fabian},
  {Farrah}, {Fryer}, {Gotthelf}, {Grindlay}, {Helfand}, {Krivonos}, {Meier},
  {Miller}, {Natalucci}, {Ogle}, {Ofek}, {Ptak}, {Reynolds}, {Rigby},
  {Tagliaferri}, {Thorsett}, {Treister}, \& {Urry}}]{Harrison2013}
{Harrison}, F.~A., {Craig}, W.~W., {Christensen}, F.~E., {et~al.} 2013, \apj,
  770, 103, \dodoi{10.1088/0004-637X/770/2/103}

\bibitem[{{Hopkins} \& {Quataert}(2010)}]{Hopkins2010}
{Hopkins}, P.~F., \& {Quataert}, E. 2010, \mnras, 407, 1529,
  \dodoi{10.1111/j.1365-2966.2010.17064.x}

\bibitem[{{Izumi} {et~al.}(2023){Izumi}, {Wada}, {Imanishi}, {Nakanishi},
  {Kohno}, {Kudoh}, {Kawamuro}, {Baba}, {Matsumoto}, {Fujita}, \&
  {Tristram}}]{Izumi2023}
{Izumi}, T., {Wada}, K., {Imanishi}, M., {et~al.} 2023, arXiv e-prints,
  arXiv:2305.03993, \dodoi{10.48550/arXiv.2305.03993}

\bibitem[{{Jansen} {et~al.}(2001){Jansen}, {Lumb}, {Altieri}, {Clavel}, {Ehle},
  {Erd}, {Gabriel}, {Guainazzi}, {Gondoin}, {Much}, {Munoz}, {Santos},
  {Schartel}, {Texier}, \& {Vacanti}}]{Jansen2001}
{Jansen}, F., {Lumb}, D., {Altieri}, B., {et~al.} 2001, \aap, 365, L1,
  \dodoi{10.1051/0004-6361:20000036}

\bibitem[{{Kimura} {et~al.}(2021){Kimura}, {Murase}, \&
  {Bartos}}]{Kimura2021_BubblesBHMs}
{Kimura}, S.~S., {Murase}, K., \& {Bartos}, I. 2021, \apj, 916, 111,
  \dodoi{10.3847/1538-4357/ac0535}

\bibitem[{{Kinugawa} {et~al.}(2014){Kinugawa}, {Inayoshi}, {Hotokezaka},
  {Nakauchi}, \& T.}]{Kinugawa14}
{Kinugawa}, T., {Inayoshi}, K., {Hotokezaka}, K., {Nakauchi}, D., \& T., N.
  2014, \mnras, 442, 2963.
\newblock \url{http://adsabs.harvard.edu/abs/2014MNRAS.442.2963K}

\bibitem[{{Kitaki} {et~al.}(2021){Kitaki}, {Mineshige}, {Ohsuga}, \&
  {Kawashima}}]{Kitaki2021}
{Kitaki}, T., {Mineshige}, S., {Ohsuga}, K., \& {Kawashima}, T. 2021, \pasj,
  73, 450, \dodoi{10.1093/pasj/psab011}

\bibitem[{{Kormendy} \& {Ho}(2013)}]{Kormendy2013}
{Kormendy}, J., \& {Ho}, L.~C. 2013, \araa, 51, 511,
  \dodoi{10.1146/annurev-astro-082708-101811}

\bibitem[{{Kumamoto} {et~al.}(2018){Kumamoto}, {Fujii}, \&
  {Tanikawa}}]{Kumamoto18}
{Kumamoto}, J., {Fujii}, M.~S., \& {Tanikawa}, A. 2018, arXiv e-prints.
\newblock \doarXiv{1811.06726}

\bibitem[{{Lazzati} {et~al.}(2022){Lazzati}, {Soares}, \&
  {Perna}}]{Lazzati2022}
{Lazzati}, D., {Soares}, G., \& {Perna}, R. 2022, \apjl, 938, L18,
  \dodoi{10.3847/2041-8213/ac98ad}

\bibitem[{{Levin} \& {Beloborodov}(2003)}]{Levin2003}
{Levin}, Y., \& {Beloborodov}, A.~M. 2003, \apjl, 590, L33,
  \dodoi{10.1086/376675}

\bibitem[{{Li} {et~al.}(2023){Li}, {Dempsey}, {Li}, {Lai}, \&
  {Li}}]{LiJiaru2023}
{Li}, J., {Dempsey}, A.~M., {Li}, H., {Lai}, D., \& {Li}, S. 2023, \apjl, 944,
  L42, \dodoi{10.3847/2041-8213/acb934}

\bibitem[{{Li} {et~al.}(2022){Li}, {Lai}, \& {Rodet}}]{LiJiaru2022}
{Li}, J., {Lai}, D., \& {Rodet}, L. 2022, \apj, 934, 154,
  \dodoi{10.3847/1538-4357/ac7c0d}

\bibitem[{{Li} {et~al.}(2021){Li}, {Dempsey}, {Li}, {Li}, \&
  {Li}}]{Li2021_BinaryEv}
{Li}, Y.-P., {Dempsey}, A.~M., {Li}, S., {Li}, H., \& {Li}, J. 2021, \apj, 911,
  124, \dodoi{10.3847/1538-4357/abed48}

\bibitem[{{Liska} {et~al.}(2018){Liska}, {Hesp}, {Tchekhovskoy}, {Ingram}, {van
  der Klis}, \& {Markoff}}]{Liska2018}
{Liska}, M., {Hesp}, C., {Tchekhovskoy}, A., {et~al.} 2018, \mnras, 474, L81,
  \dodoi{10.1093/mnrasl/slx174}

\bibitem[{{Liska} {et~al.}(2021){Liska}, {Hesp}, {Tchekhovskoy}, {Ingram}, {van
  der Klis}, {Markoff}, \& {Van Moer}}]{Liska2021}
---. 2021, \mnras, 507, 983, \dodoi{10.1093/mnras/staa099}

\bibitem[{{Martinez} {et~al.}(2022){Martinez}, {Rodriguez}, \&
  {Fragione}}]{Martinez2022}
{Martinez}, M. A.~S., {Rodriguez}, C.~L., \& {Fragione}, G. 2022, \apj, 937,
  78, \dodoi{10.3847/1538-4357/ac8d55}

\bibitem[{{McKernan} {et~al.}(2018){McKernan}, {Ford}, {Bellovary}, {Leigh},
  {Haiman}, {Kocsis}, {Lyra}, {Mac Low}, {Metzger}, {O'Dowd}, {Endlich}, \&
  {Rosen}}]{McKernan17}
{McKernan}, B., {Ford}, K.~E.~S., {Bellovary}, J., {et~al.} 2018, \apj, 866,
  66, \dodoi{10.3847/1538-4357/aadae5}

\bibitem[{{McKernan} {et~al.}(2019){McKernan}, {Ford}, {Bartos}, {Graham},
  {Lyra}, {Marka}, {Marka}, {Ross}, {Stern}, \& {Yang}}]{McKernan2019_EM}
{McKernan}, B., {Ford}, K.~E.~S., {Bartos}, I., {et~al.} 2019, \apjl, 884, L50,
  \dodoi{10.3847/2041-8213/ab4886}

\bibitem[{{Michaely} \& {Perets}(2019)}]{Michaely19}
{Michaely}, E., \& {Perets}, H.~B. 2019, \apjl, 887, L36,
  \dodoi{10.3847/2041-8213/ab5b9b}

\bibitem[{{Milosavljevi{\'c}} \& {Loeb}(2004)}]{Milosavljevic2004}
{Milosavljevi{\'c}}, M., \& {Loeb}, A. 2004, \apjl, 604, L45,
  \dodoi{10.1086/383467}

\bibitem[{{Miralda-Escud{\'e}} \& {Kollmeier}(2005)}]{Miralda-Escude2005}
{Miralda-Escud{\'e}}, J., \& {Kollmeier}, J.~A. 2005, \apj, 619, 30,
  \dodoi{10.1086/426467}

\bibitem[{{Morag} {et~al.}(2023){Morag}, {Sapir}, \& {Waxman}}]{Morag23}
{Morag}, J., {Sapir}, N., \& {Waxman}, E. 2023, \mnras, 522, 2764,
  \dodoi{10.1093/mnras/stad899}

\bibitem[{{Morton} {et~al.}(2023){Morton}, {Rinaldi}, {Torres-Orjuela},
  {Derdzinski}, {Vaccaro}, \& {Del Pozzo}}]{Morton2023}
{Morton}, S., {Rinaldi}, S., {Torres-Orjuela}, A., {et~al.} 2023, arXiv
  e-prints, arXiv:2310.16025.
\newblock \doarXiv{2310.16025}

\bibitem[{{Nakar} \& {Piran}(2017)}]{Nakar2017}
{Nakar}, E., \& {Piran}, T. 2017, \apj, 834, 28,
  \dodoi{10.3847/1538-4357/834/1/28}

\bibitem[{{Nakar} \& {Sari}(2010)}]{Nakar2010}
{Nakar}, E., \& {Sari}, R. 2010, \apj, 725, 904,
  \dodoi{10.1088/0004-637X/725/1/904}

\bibitem[{{Narayan} {et~al.}(2021){Narayan}, {Chael}, {Chatterjee}, {Ricarte},
  \& {Curd}}]{Narayan2021}
{Narayan}, R., {Chael}, A., {Chatterjee}, K., {Ricarte}, A., \& {Curd}, B.
  2021, arXiv e-prints, arXiv:2108.12380.
\newblock \doarXiv{2108.12380}

\bibitem[{{O'Leary} {et~al.}(2016){O'Leary}, {Meiron}, \& {Kocsis}}]{OLeary16}
{O'Leary}, R.~M., {Meiron}, Y., \& {Kocsis}, B. 2016, \apjl, 824, L12,
  \dodoi{10.3847/2041-8205/824/1/L12}

\bibitem[{{Ostriker}(1983)}]{Ostriker1983}
{Ostriker}, J.~P. 1983, \apj, 273, 99, \dodoi{10.1086/161351}

\bibitem[{{Palmese} {et~al.}(2021){Palmese}, {Fishbach}, {Burke}, {Annis}, \&
  {Liu}}]{Palmese2021}
{Palmese}, A., {Fishbach}, M., {Burke}, C.~J., {Annis}, J., \& {Liu}, X. 2021,
  \apjl, 914, L34, \dodoi{10.3847/2041-8213/ac0883}

\bibitem[{{Panaitescu} \& {Kumar}(2001)}]{Panaitescu2001}
{Panaitescu}, A., \& {Kumar}, P. 2001, \apjl, 560, L49, \dodoi{10.1086/324061}

\bibitem[{{Perna} {et~al.}(2021{\natexlab{a}}){Perna}, {Lazzati}, \&
  {Cantiello}}]{Perna2021_GRBs}
{Perna}, R., {Lazzati}, D., \& {Cantiello}, M. 2021{\natexlab{a}}, \apjl, 906,
  L7, \dodoi{10.3847/2041-8213/abd319}

\bibitem[{{Perna} {et~al.}(2021{\natexlab{b}}){Perna}, {Tagawa}, {Haiman}, \&
  {Bartos}}]{Perna2021_AICs}
{Perna}, R., {Tagawa}, H., {Haiman}, Z., \& {Bartos}, I. 2021{\natexlab{b}},
  \apj, 915, 10, \dodoi{10.3847/1538-4357/abfdb4}

\bibitem[{{Perna} {et~al.}(2019){Perna}, {Wang}, {Farr}, {Leigh}, \&
  {Cantiello}}]{Perna2019}
{Perna}, R., {Wang}, Y.-H., {Farr}, W.~M., {Leigh}, N., \& {Cantiello}, M.
  2019, \apjl, 878, L1, \dodoi{10.3847/2041-8213/ab2336}

\bibitem[{{Piro} {et~al.}(2021){Piro}, {Haynie}, \& {Yao}}]{Piro2021}
{Piro}, A.~L., {Haynie}, A., \& {Yao}, Y. 2021, \apj, 909, 209,
  \dodoi{10.3847/1538-4357/abe2b1}

\bibitem[{{Planck Collaboration} {et~al.}(2016){Planck Collaboration}, {Ade},
  {Aghanim}, {Arnaud}, {Ashdown}, {Aumont}, {Baccigalupi}, {Banday},
  {Barreiro}, {Bartlett}, {Bartolo}, {Battaner}, {Battye}, {Benabed},
  {Beno{\^\i}t}, {Benoit-L{\'e}vy}, {Bernard}, {Bersanelli}, {Bielewicz},
  {Bock}, {Bonaldi}, {Bonavera}, {Bond}, {Borrill}, {Bouchet}, {Boulanger},
  {Bucher}, {Burigana}, {Butler}, {Calabrese}, {Cardoso}, {Catalano},
  {Challinor}, {Chamballu}, {Chary}, {Chiang}, {Chluba}, {Christensen},
  {Church}, {Clements}, {Colombi}, {Colombo}, {Combet}, {Coulais}, {Crill},
  {Curto}, {Cuttaia}, {Danese}, {Davies}, {Davis}, {de Bernardis}, {de Rosa},
  {de Zotti}, {Delabrouille}, {D{\'e}sert}, {Di Valentino}, {Dickinson},
  {Diego}, {Dolag}, {Dole}, {Donzelli}, {Dor{\'e}}, {Douspis}, {Ducout},
  {Dunkley}, {Dupac}, {Efstathiou}, {Elsner}, {En{\ss}lin}, {Eriksen},
  {Farhang}, {Fergusson}, {Finelli}, {Forni}, {Frailis}, {Fraisse},
  {Franceschi}, {Frejsel}, {Galeotta}, {Galli}, {Ganga}, {Gauthier}, {Gerbino},
  {Ghosh}, {Giard}, {Giraud-H{\'e}raud}, {Giusarma}, {Gjerl{\o}w},
  {Gonz{\'a}lez-Nuevo}, {G{\'o}rski}, {Gratton}, {Gregorio}, {Gruppuso},
  {Gudmundsson}, {Hamann}, {Hansen}, {Hanson}, {Harrison}, {Helou},
  {Henrot-Versill{\'e}}, {Hern{\'a}ndez-Monteagudo}, {Herranz}, {Hildebrandt},
  {Hivon}, {Hobson}, {Holmes}, {Hornstrup}, {Hovest}, {Huang}, {Huffenberger},
  {Hurier}, {Jaffe}, {Jaffe}, {Jones}, {Juvela}, {Keih{\"a}nen}, {Keskitalo},
  {Kisner}, {Kneissl}, {Knoche}, {Knox}, {Kunz}, {Kurki-Suonio}, {Lagache},
  {L{\"a}hteenm{\"a}ki}, {Lamarre}, {Lasenby}, {Lattanzi}, {Lawrence}, {Leahy},
  {Leonardi}, {Lesgourgues}, {Levrier}, {Lewis}, {Liguori}, {Lilje},
  {Linden-V{\o}rnle}, {L{\'o}pez-Caniego}, {Lubin}, {Mac{\'\i}as-P{\'e}rez},
  {Maggio}, {Maino}, {Mandolesi}, {Mangilli}, {Marchini}, {Maris}, {Martin},
  {Martinelli}, {Mart{\'\i}nez-Gonz{\'a}lez}, {Masi}, {Matarrese}, {McGehee},
  {Meinhold}, {Melchiorri}, {Melin}, {Mendes}, {Mennella}, {Migliaccio},
  {Millea}, {Mitra}, {Miville-Desch{\^e}nes}, {Moneti}, {Montier}, {Morgante},
  {Mortlock}, {Moss}, {Munshi}, {Murphy}, {Naselsky}, {Nati}, {Natoli},
  {Netterfield}, {N{\o}rgaard-Nielsen}, {Noviello}, {Novikov}, {Novikov},
  {Oxborrow}, {Paci}, {Pagano}, {Pajot}, {Paladini}, {Paoletti}, {Partridge},
  {Pasian}, {Patanchon}, {Pearson}, {Perdereau}, {Perotto}, {Perrotta},
  {Pettorino}, {Piacentini}, {Piat}, {Pierpaoli}, {Pietrobon}, {Plaszczynski},
  {Pointecouteau}, {Polenta}, {Popa}, {Pratt}, {Pr{\'e}zeau}, {Prunet},
  {Puget}, {Rachen}, {Reach}, {Rebolo}, {Reinecke}, {Remazeilles}, {Renault},
  {Renzi}, {Ristorcelli}, {Rocha}, {Rosset}, {Rossetti}, {Roudier},
  {Rouill{\'e} d'Orfeuil}, {Rowan-Robinson}, {Rubi{\~n}o-Mart{\'\i}n},
  {Rusholme}, {Said}, {Salvatelli}, {Salvati}, {Sandri}, {Santos},
  {Savelainen}, {Savini}, {Scott}, {Seiffert}, {Serra}, {Shellard}, {Spencer},
  {Spinelli}, {Stolyarov}, {Stompor}, {Sudiwala}, {Sunyaev}, {Sutton},
  {Suur-Uski}, {Sygnet}, {Tauber}, {Terenzi}, {Toffolatti}, {Tomasi},
  {Tristram}, {Trombetti}, {Tucci}, {Tuovinen}, {T{\"u}rler}, {Umana},
  {Valenziano}, {Valiviita}, {Van Tent}, {Vielva}, {Villa}, {Wade}, {Wandelt},
  {Wehus}, {White}, {White}, {Wilkinson}, {Yvon}, {Zacchei}, \&
  {Zonca}}]{Planck2016}
{Planck Collaboration}, {Ade}, P.~A.~R., {Aghanim}, N., {et~al.} 2016, \aap,
  594, A13, \dodoi{10.1051/0004-6361/201525830}

\bibitem[{{Polko} \& {McKinney}(2017)}]{Polko2017}
{Polko}, P., \& {McKinney}, J.~C. 2017, \mnras, 464, 2660,
  \dodoi{10.1093/mnras/stw1875}

\bibitem[{{Portegies Zwart} \& {McMillan}(2000)}]{PortegiesZwart00}
{Portegies Zwart}, S.~F., \& {McMillan}, S.~L.~W. 2000, \apj, 528, L17.
\newblock \url{http://adsabs.harvard.edu/abs/2000ApJ...528L..17P}

\bibitem[{{Prasad} {et~al.}(2023){Prasad}, {Wang}, {Perna}, {Ford}, \&
  {McKernan}}]{Prasad2023}
{Prasad}, C., {Wang}, Y., {Perna}, R., {Ford}, K.~E.~S., \& {McKernan}, B.
  2023, arXiv e-prints, arXiv:2310.00020, \dodoi{10.48550/arXiv.2310.00020}

\bibitem[{{Qian} {et~al.}(2023){Qian}, {Li}, \& {Lai}}]{Qian2023}
{Qian}, K., {Li}, J., \& {Lai}, D. 2023, arXiv e-prints, arXiv:2310.12208,
  \dodoi{10.48550/arXiv.2310.12208}

\bibitem[{{Rasskazov} \& {Kocsis}(2019)}]{Rasskazov19}
{Rasskazov}, A., \& {Kocsis}, B. 2019, \apj, 881, 20,
  \dodoi{10.3847/1538-4357/ab2c74}

\bibitem[{{Ray} {et~al.}(2023){Ray}, {Lazzati}, \& {Perna}}]{Ray2023}
{Ray}, M., {Lazzati}, D., \& {Perna}, R. 2023, \mnras, 521, 4233,
  \dodoi{10.1093/mnras/stad816}

\bibitem[{{Rodriguez} {et~al.}(2016){Rodriguez}, {Chatterjee}, \&
  {Rasio}}]{Rodriguez16}
{Rodriguez}, C.~L., {Chatterjee}, S., \& {Rasio}, F.~A. 2016, Phys. Rev. D.,
  93, 084029.
\newblock \url{http://adsabs.harvard.edu/abs/2016PhRvD..93h4029R}

\bibitem[{{Rodr{\'\i}guez-Ram{\'\i}rez}
  {et~al.}(2023){Rodr{\'\i}guez-Ram{\'\i}rez}, {Bom}, {Fraga}, \&
  {Nemmen}}]{RodriguezRamirez2023}
{Rodr{\'\i}guez-Ram{\'\i}rez}, J.~C., {Bom}, C.~R., {Fraga}, B., \& {Nemmen},
  R. 2023, arXiv e-prints, arXiv:2304.10567, \dodoi{10.48550/arXiv.2304.10567}

\bibitem[{{Romero-Shaw} {et~al.}(2022){Romero-Shaw}, {Lasky}, \&
  {Thrane}}]{RomeroShaw2022}
{Romero-Shaw}, I., {Lasky}, P.~D., \& {Thrane}, E. 2022, \apj, 940, 171,
  \dodoi{10.3847/1538-4357/ac9798}

\bibitem[{{Romero-Shaw} {et~al.}(2023){Romero-Shaw}, {Gerosa}, \&
  {Loutrel}}]{RomeroShaw2023}
{Romero-Shaw}, I.~M., {Gerosa}, D., \& {Loutrel}, N. 2023, \mnras, 519, 5352,
  \dodoi{10.1093/mnras/stad031}

\bibitem[{{Romero-Shaw} {et~al.}(2020){Romero-Shaw}, {Lasky}, {Thrane}, \&
  {Calderon Bustillo}}]{Romero-Shaw20}
{Romero-Shaw}, I.~M., {Lasky}, P.~D., {Thrane}, E., \& {Calderon Bustillo}, J.
  2020, arXiv e-prints, arXiv:2009.04771.
\newblock \doarXiv{2009.04771}

\bibitem[{{Rossi} {et~al.}(2010){Rossi}, {Lodato}, {Armitage}, {Pringle}, \&
  {King}}]{Rossi2010}
{Rossi}, E.~M., {Lodato}, G., {Armitage}, P.~J., {Pringle}, J.~E., \& {King},
  A.~R. 2010, \mnras, 401, 2021, \dodoi{10.1111/j.1365-2966.2009.15802.x}

\bibitem[{{Rowan} {et~al.}(2022){Rowan}, {Boekholt}, {Kocsis}, \&
  {Haiman}}]{Rowan2022}
{Rowan}, C., {Boekholt}, T., {Kocsis}, B., \& {Haiman}, Z. 2022, arXiv
  e-prints, arXiv:2212.06133, \dodoi{10.48550/arXiv.2212.06133}

\bibitem[{{Rowan} {et~al.}(2023){Rowan}, {Whitehead}, {Boekholt}, {Kocsis}, \&
  {Haiman}}]{Rowan2023}
{Rowan}, C., {Whitehead}, H., {Boekholt}, T., {Kocsis}, B., \& {Haiman}, Z.
  2023, arXiv e-prints, arXiv:2309.14433, \dodoi{10.48550/arXiv.2309.14433}

\bibitem[{{Rozner} {et~al.}(2023){Rozner}, {Generozov}, \&
  {Perets}}]{Rozner2023}
{Rozner}, M., {Generozov}, A., \& {Perets}, H.~B. 2023, \mnras,
  \dodoi{10.1093/mnras/stad603}

\bibitem[{{Samsing} {et~al.}(2014){Samsing}, {MacLeod}, \&
  {Ramirez-Ruiz}}]{Samsing14}
{Samsing}, J., {MacLeod}, M., \& {Ramirez-Ruiz}, E. 2014, \apj, 784, 71,
  \dodoi{10.1088/0004-637X/784/1/71}

\bibitem[{{Samsing} {et~al.}(2020){Samsing}, {Bartos}, {D'Orazio}, {Haiman},
  {Kocsis}, {Leigh}, {Liu}, {Pessah}, \& {Tagawa}}]{Samsing20}
{Samsing}, J., {Bartos}, I., {D'Orazio}, D.~J., {et~al.} 2020, arXiv e-prints,
  arXiv:2010.09765.
\newblock \doarXiv{2010.09765}

\bibitem[{{Sapir} \& {Waxman}(2017)}]{Sapir2017}
{Sapir}, N., \& {Waxman}, E. 2017, \apj, 838, 130,
  \dodoi{10.3847/1538-4357/aa64df}

\bibitem[{{Scott} \& {Graham}(2013)}]{Scott13}
{Scott}, N., \& {Graham}, A.~W. 2013, \apj, 763, 76.
\newblock \url{http://adsabs.harvard.edu/abs/2013ApJ...763...76S}

\bibitem[{{Shakura} \& {Sunyaev}(1973)}]{Shakura73}
{Shakura}, N.~I., \& {Sunyaev}, R.~A. 1973, A\& A, 24, 337.
\newblock \url{http://adsabs.harvard.edu/abs/1973A%26A....24..337S}

\bibitem[{{Shvartzvald} {et~al.}(2023){Shvartzvald}, {Waxman}, {Gal-Yam},
  {Ofek}, {Ben-Ami}, {Berge}, {Kowalski}, {B{\"u}hler}, {Worm}, {Rhoads},
  {Arcavi}, {Maoz}, {Polishook}, {Stone}, {Trakhtenbrot}, {Ackermann},
  {Aharonson}, {Birnholtz}, {Chelouche}, {Guetta}, {Hallakoun}, {Horesh},
  {Kushnir}, {Mazeh}, {Nordin}, {Ofir}, {Ohm}, {Parsons}, {Pe'er}, {Perets},
  {Perdelwitz}, {Poznanski}, {Sadeh}, {Sagiv}, {Shahaf}, {Soumagnac}, {Tal-Or},
  {Van Santen}, {Zackay}, {Guttman}, {Rekhi}, {Townsend}, {Weinstein}, \&
  {Wold}}]{Shvartzvald2023}
{Shvartzvald}, Y., {Waxman}, E., {Gal-Yam}, A., {et~al.} 2023, arXiv e-prints,
  arXiv:2304.14482, \dodoi{10.48550/arXiv.2304.14482}

\bibitem[{{Silsbee} \& {Tremaine}(2017)}]{Silsbee17}
{Silsbee}, K., \& {Tremaine}, S. 2017, \apj, 836, 39.
\newblock \url{http://adsabs.harvard.edu/abs/2017ApJ...836...39S}

\bibitem[{{Spera} {et~al.}(2019){Spera}, {Mapelli}, {Giacobbo}, {Trani},
  {Bressan}, \& {Costa}}]{Spera19}
{Spera}, M., {Mapelli}, M., {Giacobbo}, N., {et~al.} 2019, \mnras, 485, 889,
  \dodoi{10.1093/mnras/stz359}

\bibitem[{{Stone} {et~al.}(2017){Stone}, {Metzger}, \& {Haiman}}]{Stone17}
{Stone}, N.~C., {Metzger}, B.~D., \& {Haiman}, Z. 2017, \mnras, 464, 946.
\newblock \url{http://adsabs.harvard.edu/abs/2017MNRAS.464..946S}

\bibitem[{{Tagawa} {et~al.}(2020{\natexlab{a}}){Tagawa}, {Haiman}, {Bartos}, \&
  {Kocsis}}]{Tagawa20b_spin}
{Tagawa}, H., {Haiman}, Z., {Bartos}, I., \& {Kocsis}, B. 2020{\natexlab{a}},
  \apj, 899, 26, \dodoi{10.3847/1538-4357/aba2cc}

\bibitem[{{Tagawa} {et~al.}(2020{\natexlab{b}}){Tagawa}, {Haiman}, \&
  {Kocsis}}]{Tagawa19}
{Tagawa}, H., {Haiman}, Z., \& {Kocsis}, B. 2020{\natexlab{b}}, \apj, 898, 25,
  \dodoi{10.3847/1538-4357/ab9b8c}

\bibitem[{{Tagawa} {et~al.}(2023{\natexlab{a}}){Tagawa}, {Kimura}, \&
  {Haiman}}]{Tagawa2023_highenergy}
{Tagawa}, H., {Kimura}, S.~S., \& {Haiman}, Z. 2023{\natexlab{a}}, \apj, 955,
  23, \dodoi{10.3847/1538-4357/ace71d}

\bibitem[{{Tagawa} {et~al.}(2023{\natexlab{b}}){Tagawa}, {Kimura}, {Haiman},
  {Perna}, \& {Bartos}}]{Tagawa2023}
{Tagawa}, H., {Kimura}, S.~S., {Haiman}, Z., {Perna}, R., \& {Bartos}, I.
  2023{\natexlab{b}}, \apj, 950, 13, \dodoi{10.3847/1538-4357/acc4bb}

\bibitem[{{Tagawa} {et~al.}(2023{\natexlab{c}}){Tagawa}, {Kimura}, {Haiman},
  {Perna}, \& {Bartos}}]{Tagawa2023_solitary}
---. 2023{\natexlab{c}}, \apjl, 946, L3, \dodoi{10.3847/2041-8213/acc103}

\bibitem[{{Tagawa} {et~al.}(2022){Tagawa}, {Kimura}, {Haiman}, {Perna},
  {Tanaka}, \& {Bartos}}]{Tagawa2022_BHFeedback}
{Tagawa}, H., {Kimura}, S.~S., {Haiman}, Z., {et~al.} 2022, \apj, 927, 41,
  \dodoi{10.3847/1538-4357/ac45f8}

\bibitem[{{Tagawa} {et~al.}(2021{\natexlab{a}}){Tagawa}, {Kocsis}, {Haiman},
  {Bartos}, {Omukai}, \& {Samsing}}]{Tagawa20_MassGap}
{Tagawa}, H., {Kocsis}, B., {Haiman}, Z., {et~al.} 2021{\natexlab{a}}, \apj,
  908, 194, \dodoi{10.3847/1538-4357/abd555}

\bibitem[{{Tagawa} {et~al.}(2021{\natexlab{b}}){Tagawa}, {Kocsis}, {Haiman},
  {Bartos}, {Omukai}, \& {Samsing}}]{Tagawa20_ecc}
---. 2021{\natexlab{b}}, \apjl, 907, L20, \dodoi{10.3847/2041-8213/abd4d3}

\bibitem[{{Tagawa} {et~al.}(2018){Tagawa}, {Kocsis}, \& {Saitoh}}]{Tagawa18}
{Tagawa}, H., {Kocsis}, B., \& {Saitoh}, R.~T. 2018, Phys. Rev. Lett., 120,
  261101.
\newblock \url{http://adsabs.harvard.edu/abs/2018PhRvL.120z1101T}

\bibitem[{{Tanigawa} \& {Tanaka}(2016)}]{Tanigawa2016}
{Tanigawa}, T., \& {Tanaka}, H. 2016, \apj, 823, 48,
  \dodoi{10.3847/0004-637X/823/1/48}

\bibitem[{{Tanigawa} \& {Watanabe}(2002)}]{Tanigawa2002}
{Tanigawa}, T., \& {Watanabe}, S.-i. 2002, \apj, 580, 506,
  \dodoi{10.1086/343069}

\bibitem[{{Tanikawa} {et~al.}(2022){Tanikawa}, {Yoshida}, {Kinugawa}, {Trani},
  {Hosokawa}, {Susa}, \& {Omukai}}]{Tanikawa2022}
{Tanikawa}, A., {Yoshida}, T., {Kinugawa}, T., {et~al.} 2022, \apj, 926, 83,
  \dodoi{10.3847/1538-4357/ac4247}

\bibitem[{{Tchekhovskoy} {et~al.}(2010){Tchekhovskoy}, {Narayan}, \&
  {McKinney}}]{Tchekhovskoy2010}
{Tchekhovskoy}, A., {Narayan}, R., \& {McKinney}, J.~C. 2010, \apj, 711, 50,
  \dodoi{10.1088/0004-637X/711/1/50}

\bibitem[{{The LIGO Scientific Collaboration} {et~al.}(2021){The LIGO
  Scientific Collaboration}, {the Virgo Collaboration}, {the KAGRA
  Collaboration}, {Abbott}, {Abbott}, {Acernese}, {Ackley}, {Adams},
  {Adhikari}, {Adhikari}, {Adya}, {Affeldt}, {Agarwal}, {Agathos}, {Agatsuma},
  {Aggarwal}, {Aguiar}, {Aiello}, {Ain}, {Ajith}, {Akcay}, {Akutsu},
  {Albanesi}, {Allocca}, {Altin}, {Amato}, {Anand}, {Anand}, {Ananyeva},
  {Anderson}, {Anderson}, \& {Ando}}]{Abbott21_GWTC3}
{The LIGO Scientific Collaboration}, {the Virgo Collaboration}, {the KAGRA
  Collaboration}, {et~al.} 2021, arXiv e-prints, arXiv:2111.03606.
\newblock \doarXiv{2111.03606}

\bibitem[{{Thompson} {et~al.}(2005){Thompson}, {Quataert}, \&
  {Murray}}]{Thompson05}
{Thompson}, T.~A., {Quataert}, E., \& {Murray}, N. 2005, \apj, 630, 167.
\newblock \url{http://adsabs.harvard.edu/abs/2005ApJ...630..167T}

\bibitem[{{Wang} {et~al.}(2021{\natexlab{a}}){Wang}, {Liu}, {Ho}, \&
  {Du}}]{Wang2021_TZW}
{Wang}, J.-M., {Liu}, J.-R., {Ho}, L.~C., \& {Du}, P. 2021{\natexlab{a}},
  \apjl, 911, L14, \dodoi{10.3847/2041-8213/abee81}

\bibitem[{{Wang} {et~al.}(2021{\natexlab{b}}){Wang}, {Liu}, {Ho}, {Li}, \&
  {Du}}]{Wang2021b}
{Wang}, J.-M., {Liu}, J.-R., {Ho}, L.~C., {Li}, Y.-R., \& {Du}, P.
  2021{\natexlab{b}}, \apjl, 916, L17, \dodoi{10.3847/2041-8213/ac0b46}

\bibitem[{{Wang} {et~al.}(2023){Wang}, {Zhu}, \& {Lin}}]{Wang2023_capture}
{Wang}, Y., {Zhu}, Z., \& {Lin}, D. N.~C. 2023, arXiv e-prints,
  arXiv:2308.09129, \dodoi{10.48550/arXiv.2308.09129}

\bibitem[{{Wang} {et~al.}(2022){Wang}, {Lazzati}, \&
  {Perna}}]{Wang2022_Afterglow}
{Wang}, Y.-H., {Lazzati}, D., \& {Perna}, R. 2022, \mnras, 516, 5935,
  \dodoi{10.1093/mnras/stac1968}

\bibitem[{{Whitehead} {et~al.}(2023){Whitehead}, {Rowan}, {Boekholt}, \&
  {Kocsis}}]{Whitehead2023}
{Whitehead}, H., {Rowan}, C., {Boekholt}, T., \& {Kocsis}, B. 2023, arXiv
  e-prints, arXiv:2309.11561, \dodoi{10.48550/arXiv.2309.11561}

\bibitem[{{Xin} {et~al.}(2023){Xin}, {Haiman}, {Perna}, {Wang}, \&
  {Ryu}}]{Xin2023}
{Xin}, C., {Haiman}, Z., {Perna}, R., {Wang}, Y., \& {Ryu}, T. 2023, arXiv
  e-prints, arXiv:2303.12846, \dodoi{10.48550/arXiv.2303.12846}

\bibitem[{{Yang} {et~al.}(2021){Yang}, {Bartos}, {Fragione}, {Haiman},
  {Kowalski}, {Marka}, {Perna}, \& {Tagawa}}]{Yang2021_TDE}
{Yang}, Y., {Bartos}, I., {Fragione}, G., {et~al.} 2021, arXiv e-prints,
  arXiv:2105.02342.
\newblock \doarXiv{2105.02342}

\bibitem[{{Yonetoku} {et~al.}(2020){Yonetoku}, {Mihara}, {Doi}, {Sakamoto},
  {Tsumura}, {Ioka}, {Amaya}, {Arimoto}, {Enoto}, {Fujii}, {Goto}, {Gunji},
  {Hiraga}, {Ikeda}, {Kawai}, {Kurosawa}, {Li}, {Maeda}, {Mitsuishi},
  {Murakami}, {Nakagawa}, {Ogino}, {Ohno}, {Sawano}, {Sei}, {Serino}, {Sugita},
  {Tamagawa}, {Tamura}, {Tanaka}, {Tanimori}, {Tashiro}, {Tomida}, {Wang},
  {Yamaguchi}, {Yamamoto}, {Yamaoka}, {Yamauchi}, {Yatsu}, {Yoshida}, {Yuhi},
  {Akitaya}, {Fukui}, {Ita}, {Kaneda}, {Kawabata}, {Kawata}, {Kurimata},
  {Matsumoto}, {Matsuura}, {Miyasaka}, {Motohara}, {Narita}, {Noda}, {Ohashi},
  {Okita}, {Sano}, {Tanaka}, {Urata}, {Wada}, {Yamaguchi}, {Yanagisawa},
  {Yoshida}, {Asano}, {Inayoshi}, {Inoue}, {Ito}, {Izumiura}, {Kawanaka},
  {Kinugawa}, {Kisaka}, {Kiuchi}, {Matsumoto}, {Mizuta}, {Murase}, {Nagakura},
  {Nagataki}, {Nakada}, {Nakamura}, {Niino}, {Suwa}, {Takahashi}, {Tanaka},
  {Toma}, {Totani}, {Yamazaki}, \& {Yokoyama}}]{Yonetoku2020c}
{Yonetoku}, D., {Mihara}, T., {Doi}, A., {et~al.} 2020, in Society of
  Photo-Optical Instrumentation Engineers (SPIE) Conference Series, Vol. 11444,
  Society of Photo-Optical Instrumentation Engineers (SPIE) Conference Series,
  114442Z, \dodoi{10.1117/12.2560603}

\bibitem[{{Yuan} {et~al.}(2021){Yuan}, {Murase}, {Guetta}, {Pe'er}, {Bartos},
  \& {M{\'e}sz{\'a}ros}}]{Yuan2021}
{Yuan}, C., {Murase}, K., {Guetta}, D., {et~al.} 2021, arXiv e-prints,
  arXiv:2112.07653.
\newblock \doarXiv{2112.07653}

\bibitem[{{Yuan} \& {Narayan}(2014)}]{Yuan2014}
{Yuan}, F., \& {Narayan}, R. 2014, \araa, 52, 529,
  \dodoi{10.1146/annurev-astro-082812-141003}

\bibitem[{{Yuan} {et~al.}(2015){Yuan}, {Zhang}, {Feng}, {Zhang}, {Ling},
  {Zhao}, {Deng}, {Qiu}, {Osborne}, {O'Brien}, {Willingale}, {Lapington},
  {Fraser}, \& {the Einstein Probe team}}]{Yuan2015}
{Yuan}, W., {Zhang}, C., {Feng}, H., {et~al.} 2015, arXiv e-prints,
  arXiv:1506.07735.
\newblock \doarXiv{1506.07735}

\bibitem[{{Zhou} \& {Wang}(2023)}]{Zhou2023}
{Zhou}, Z.-H., \& {Wang}, K. 2023, arXiv e-prints, arXiv:2310.15832.
\newblock \doarXiv{2310.15832}

\bibitem[{{Zhu}(2023)}]{Zhu2023_neutrino}
{Zhu}, J.-P. 2023, arXiv e-prints, arXiv:2310.14255.
\newblock \doarXiv{2310.14255}

\bibitem[{{Zhu} {et~al.}(2021{\natexlab{a}}){Zhu}, {Wang}, \&
  {Zhang}}]{Zhu2021_Neutrino}
{Zhu}, J.-P., {Wang}, K., \& {Zhang}, B. 2021{\natexlab{a}}, \apjl, 917, L28,
  \dodoi{10.3847/2041-8213/ac1a17}

\bibitem[{{Zhu} {et~al.}(2021{\natexlab{b}}){Zhu}, {Yang}, {Zhang}, {Liu},
  {Yu}, \& {Gao}}]{Zhu2021_WD_AIC}
{Zhu}, J.-P., {Yang}, Y.-P., {Zhang}, B., {et~al.} 2021{\natexlab{b}}, \apjl,
  914, L19, \dodoi{10.3847/2041-8213/abff5a}

\bibitem[{{Zhu} {et~al.}(2021{\natexlab{c}}){Zhu}, {Zhang}, {Yu}, \&
  {Gao}}]{Zhu2021_Cocoon_NSMs}
{Zhu}, J.-P., {Zhang}, B., {Yu}, Y.-W., \& {Gao}, H. 2021{\natexlab{c}}, \apjl,
  906, L11, \dodoi{10.3847/2041-8213/abd412}

\end{thebibliography}

\end{document}